\documentclass{aa}  
\usepackage{graphicx}   
\usepackage{amsmath}    

\usepackage{chemformula}
\usepackage{caption}
\usepackage{subcaption}

\usepackage{txfonts}
\usepackage{hyperref}
\usepackage{lscape}

\begin{document}

   \title{The GUAPOS project}
 \subtitle{VI. The chemical inventory of  shocked gas }

   \titlerunning{The GUAPOS project. VI. Chemical inventory of shocked gas}
   \authorrunning{Á.~López-Gallifa et al.}

   \author{\'{A}.~L\'{o}pez-Gallifa\inst{1,2}, 
    V. M.~Rivilla\inst{1},
    M. T. Beltrán\inst{3},
    L. Colzi\inst{1},
    F. Fontani\inst{3},
    Á. Sánchez-Monge\inst{4,5},
    C. Mininni\inst{6},
    R. Cesaroni\inst{3},
    I. Jiménez-Serra\inst{1},  
    S. Viti\inst{7,8}
    and A. Lorenzani\inst{3}
    }

   \institute{Centro de Astrobiología (CAB), CSIC-INTA, Ctra. de Ajalvir, km. 4,             Torrejón de Ardoz, E-28850 Madrid, Spain
      \and
   Departamento de Física de la Tierra y Astrofísica, Facultad de Ciencias Físicas, Universidad Complutense de Madrid, 28040 Madrid, Spain 
         \and
 INAF-Osservatorio Astrofisico di Arcetri, Largo E. Fermi 5, I-50125, Florence, Italy
 \and
  Institut de Ci\`encies de l'Espai (ICE, CSIC), Carrer de Can Magrans, s/n, E-08193 Bellaterra, Barcelona, Spain
 \and
 Institut d'Estudis Espacials de Catalunya (IEEC), Barcelona, Spain
  \and
 INAF-IAPS, via del Fosso del Cavaliere 100, I-00133 Roma, Italy
   \and
 Leiden Observatory, Leiden University, Huygens Laboratory, Niels Bohrweg 2, NL-2333 CA Leiden, The Netherlands
   \and
 Department of Physics and Astronomy, University College London, Gower Street, WC1E 6BT, London, UK
}

   \date{}

\abstract
{The study of the chemical composition of star-forming regions is key to understanding the chemical ingredients available during the formation of planetary systems. Because the chemical inventory of interstellar dust grains in the prestellar phases might be altered by protostellar warming, an alternative to inferring the chemical composition of the grains might be to observe regions that are affected by shocks associated with molecular outflows. These shocks are able to desorb the molecules and might produce less chemical processing because the timescales are shorter. We present a detailed study of the chemical reservoir of a shocked region located in the G31.41+0.31 protocluster using data from the G31.41+0.31 Unbiased ALMA sPectral Observational Survey (GUAPOS). We report the detection of 30 molecular species (plus 18 isotopologs) and derived the column densities. We compared the molecular ratios in the shocked region with those derived toward the hot core of G31.41+0.31. They are poorly correlated, with the exception of N-bearing species. Our results confirm observationally that a different level of chemical alteration is present in hot cores and in shocks. While the former likely alter the molecular ratios by thermal processing during longer timescales, the latter might represent freshly desorbed material that constitutes a better proxy of the composition of the ice mantle.
The similarity of the molecular ratios of the N-bearing species in the G31.41+0.31 shock and the hot core suggests that these species are predominantly formed at early evolutionary stages.
Interestingly, the abundances in the G31.41+0.31 shock are better correlated with other shock-dominated regions (two protostellar outflows and a molecular cloud in the Galactic center). This suggests that gas-phase chemistry after shock-induced ejection from grains is negligible and that the composition of the ice mantle is similar regardless of the Galactic environment.

}

   \keywords{Astrochemistry - Line: identification - ISM: molecules - ISM: individual object: G31.41+0.31 - Stars: formation}

   \maketitle

\section{Introduction}

The high-temperature gas environment surrounding newly born massive stars, usually known as a hot core, exhibits a rich chemical reservoir. A significant fraction of interstellar molecules was discovered toward hot cores, such as Sgr B2(N), Orion KL, or NGC 6334I (e.g., \citealt{Belloche2008, Tercero2013, Fried2024}, respectively). This shows that
hot cores play a crucial role in astrochemistry that allows us to unveil the molecular inventory of the interstellar medium (ISM). 
There is increasing evidence that the chemical content of stellar nurseries is at least partially inherited in the later stages of star and planet formation (e.g., \citealt{Loomis2018, Bianchi2019, Booth2021, vanderMarel2021, Ceccarelli2023, Tobin2023}). 
Our Sun was born in a dense stellar cluster environment that included massive stars (\citealt{Adams2001, Malmberg2007, Portegies_Zwart2009, Adams2010, Dukes2012, Korschinek2020, PfalznerVincke2020, Brinkman2021}), and the study of the molecular reservoir of high-mass star-forming regions can therefore give us useful insights into the molecules that were available when our Solar System was born. 

The chemical richness of hot cores is certainly favored by the heating produced by the central protostars, which gradually increases the temperature on timescales of $10^4$-$10^6$ yr \citep{Kurtz2000_libro, Garrod2013a} to values of $T$ $>$ 100 K (\citealt{Choudhury2015}). The availability of this thermal budget enhances the mobility of the molecules on the grain surfaces, which can react with other species and form more complex species. It also allows the desorption into the gas phase (e.g., \citealt{Garrod2022}).
Moreover, some gas-phase reactions become efficient at these high temperatures, which can contribute to a further increase of the molecular complexity (\citealt{Charnley1992, Skouteris2018, Gieser2019, Barger2020}).
As a result of this thermal processing, the gas-phase abundances of the hot cores might significantly differ from the initial abundances on the surface of dust grains. An alternative to estimating molecular abundances on the icy grain mantles is provided today by the James Webb Space Telescope (JWST), which has the potential to identify  vibrational transitions of molecules directly on the grain surfaces in absorption against bright background sources \citep{McClure2023, Slavicinska2025}. However, the limited sensitivity of JWST and the significant spectral feature blending at infrared wavelengths prevent us from reaching a complete census of the chemical feedstock present on the grains.

Observations of gas that is desorbed by the action of shocks therefore appears to be a promising alternative. Molecular outflows powered by protostars can produce C-type low-velocity shocks that can desorb the molecular content of the icy mantles \citep{Gordard2019}. The shocked regions are usually located far away from the protostars and were therefore not heated directly. Moreover, the timescales of these shocks are significantly shorter ($10^2$-$10^3$ yr, \citealt{Jimenez-Serra2008, Burkhardt2019}) than those of protostellar heating, and thus, the possible effect of gas-phase chemistry is expected to be weaker \citep{Requena-Torres2006}. Altogether, shocked regions might offer a better proxy of the chemical budget of ices than hot cores.

We have analyzed the rich molecular reservoir of a shock-dominated region in the G31.41+0.31 protostellar cluster. This high-mass star-forming region (hereafter G31.41 core) is located at a distance of 3.75 kpc \citep{Immer2019} and has a mass of 70 M$_\odot$ \citep{Cesaroni2019}. It contains at least four massive protostars (\citealt{Beltran2021}) and harbors a well-studied hot core that is one of the most chemically rich sources in the Galaxy (e.g., \citealt{Beltran2009, Rivilla2017, Suzuki2023}).
In the past years, the project G31.41+0.31 Unbiased ALMA sPectral Observational Survey (GUAPOS; \citealt{Mininni2020}) studied the molecular inventory of this star-forming region in detail: isomers of C$_2$H$_4$O$_2$ (Paper I; \citealt{Mininni2020}), peptide-like bond molecules (Paper II; \cite{Colzi2021}, oxygen- and nitrogen-bearing complex organic molecules (COMs; Paper III; \citealt{Mininni2023}), and phosphorus-bearing molecules as shock tracers (Paper IV; \citealt{Fontani2024}). Moreover, in \cite{Lopez-Gallifa2024} (Paper V), we compared the whole chemical reservoir of the G31.41 core with that of other interstellar sources and Solar System objects representative of different evolutionary stages of star and planet formation.

\begin{figure*}[ht]

\includegraphics[scale=0.39]{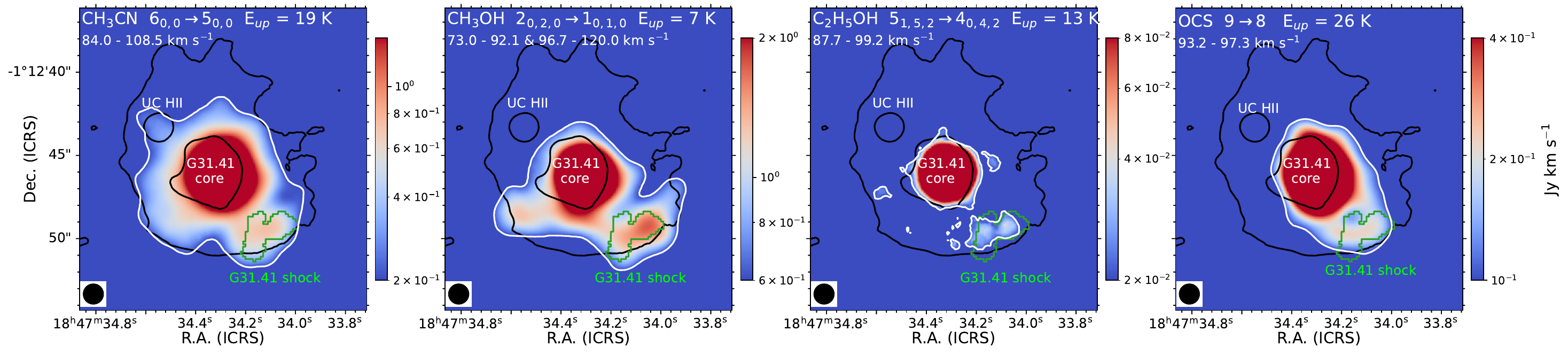}
\includegraphics[scale=0.39]{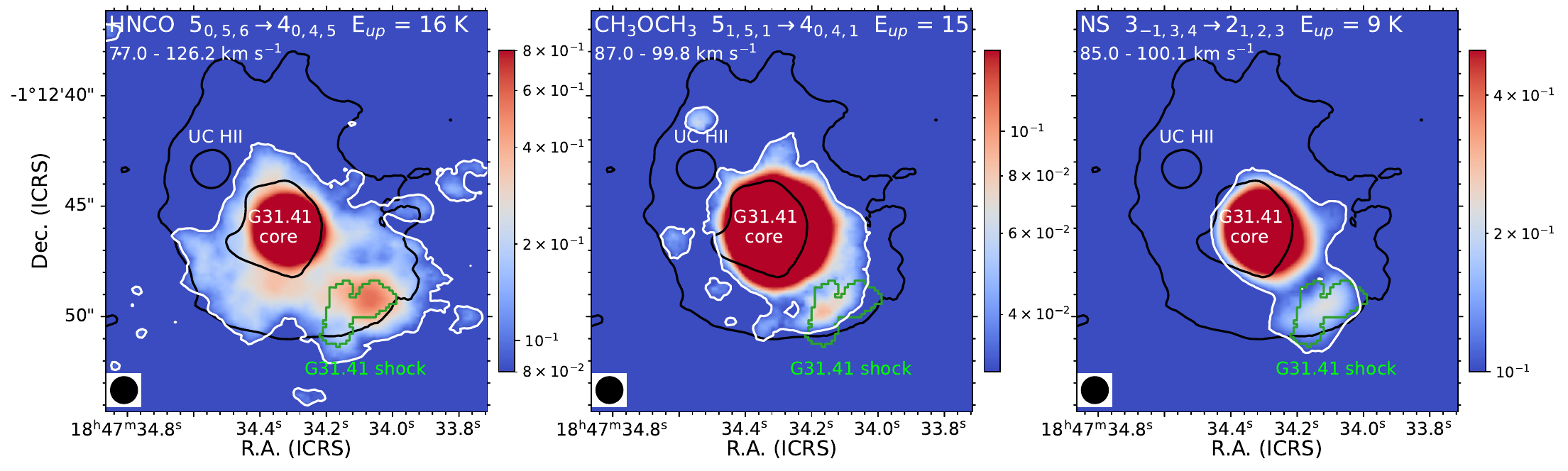}
\caption{Spatial distribution toward the G31.41 star-forming region of the molecular emission of CH$_3$CN, the broad component of CH$_3$OH, C$_2$H$_5$OH, the narrow component of OCS (upper panels from left to right), and HNCO, CH$_3$OCH$_3$, and NS (lower panels from left to right). We list the molecule name, the molecular transition, the energy of the upper level, and the velocity range in all panels in white. In addition, the G31.41 shock region according to region 2 of Fig. 1 in \cite{Fontani2024} is shown in green. The ultracompact HII region and the hot core of G31.41 are also labeled. The 3 and 20 $\sigma$ contour levels of the 3mm continuum map at 98.5 GHz from \cite{Mininni2020} are plotted in black. The minimum flux level of the color bar of each panel is indicated by white contours. 
}
\label{fig:mols_intro_shock_region}
\end{figure*}

The protostars embedded in the G31.41 core power several molecular outflows (\citealt{Beltran2022a}) that produce shocked gas in their surroundings. Southeast of the hot core, at $\sim$19000 au from the core center, observations of well-known shock tracers (sulfur- and phosphorus-bearing species) have revealed a bright shock (green contour in Fig. \ref{fig:mols_intro_shock_region}; and position 2 in Fig. 1 of \citealt{Fontani2024}). This region (hereafter G31.41 shock) offers a unique opportunity to compare the molecular abundances of the shocked gas with those of the neighboring hot core (\citealt{Mininni2020, Colzi2021, Mininni2023, Lopez-Gallifa2024}). This analysis allows us to directly compare the effect of shocks and protostellar heating on the molecular abundances.

This article is organized as follows: In Sect. \ref{sec:observations} we describe the ALMA observations, in Sect. \ref{sec:data analysis} we explain the analysis procedure we used to study the chemical reservoir of G31.41 shock, in Sect. \ref{sec:comparison} we compare the molecular inventory of the G31.41 shock position with that of the hot core and of three shock-dominated regions, in Sect. \ref{sec:discuss} we discuss the results, and in Sect. \ref{sec:conclusions} we summarize the main results and implications of this work.

\section{ALMA observations of G31.41+0.31}
\label{sec:observations}

The observations toward G31.41+0.31 were carried out with ALMA in Cycle 5 as part of project 2017.1.00501.S (PI: M. T. Beltrán). The full ALMA band 3 (86.06-115.91 GHz) was covered, which provides an unbiased spectral survey. The phase center was $\alpha_{\mathrm{ICRS}} = 18^{\mathrm{h}}45^{\mathrm{m}}34^{\mathrm{s}}$ and $\delta_{\mathrm{ICRS}} = - 0.1^{\circ}12'45''$. The frequency resolution was 0.49 MHz, which is equivalent to a velocity resolution of $\sim$ 1.6 km s$^{-1}$ at 90 GHz. The common restoring beam created for the datacubes of the whole survey was 1.2$''$($\sim$ 4500 au). The uncertainties in the flux calibration for ALMA band 3 are $\sim$5$\%$ according to quality assessment 2 reports, which agrees well with the flux uncertainties reported by \cite{Bonato2018}.  
The data were calibrated and imaged with Common Astronomy Software Applications package (CASA\footnote{https://casa.nrao.edu}, \citealt{McMullin2007}). Further details are given in Paper I \citep{Mininni2020}.

\begin{figure*}
\centering
\includegraphics[scale=0.55]{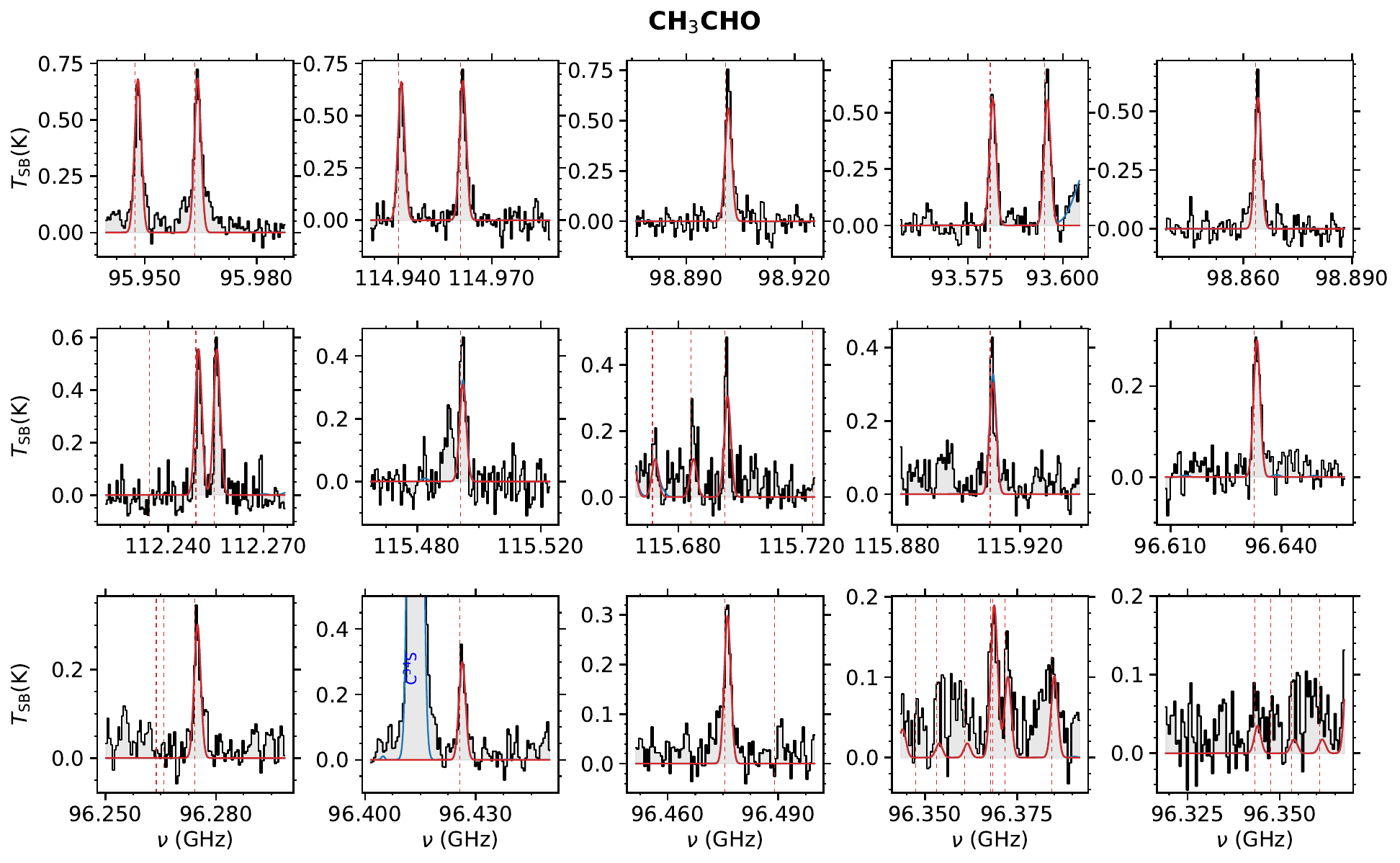}
\vspace{-2mm}
\caption{Transitions of CH$_3$CHO detected toward the G31.41 shock position. The black histogram and its gray shadow represent the observed spectrum. The red curve shows the best LTE fit of the individual species, and the blue curve shows the cumulative fit considering all detected species. The dashed red lines indicate the frequency of the transitions we fit. The plots are sorted by decreasing line intensity of the transitions.}
\label{fig:CH3CHO}
\end{figure*}

\section{Data analysis}
\label{sec:data analysis}

\subsection{Spatial distribution of the molecular emission}

To map the molecular emission of the different species, we subtracted the continuum in the GUAPOS datacubes using the code STATCONT\footnote{https://hera.ph1.uni-koeln.de/~sanchez/statcont} (\citealt{Sanchez-Monge2018}), which is a Python tool that applies a statistical analysis to infer the continuum level and produces continuum-subtracted spectral cubes. Then, we generated line integrated-intensity maps of the most unblended lines using the software Madrid Data Cube Analysis (MADCUBA\footnote{MADCUBA is developed at the Centro de Astrobiología (CAB) in Madrid, free access: https://cab.inta-csic.es/madcuba/}; \citealt{Martin2019}).
We show in Fig. \ref{fig:mols_intro_shock_region} the resulting emission maps in the G31.41 shock of methyl cyanide (CH$_3$CN), the broad component of methanol (CH$_3$OH), ethanol (C$_2$H$_5$OH), the narrow component of carbonyl sulfide (OCS), isocyanic acid (HNCO), dimethyl ether (CH$_3$OCH$_3$), and nitrogen sulfide (NS) as examples of oxygen-, nitrogen-, and sulfur-bearing species. The integrated emission maps of the other detected molecules in the G31.41 shock (see Sect. \ref{sec:detceted_mols}) are shown in Appendix \ref{sec:Emission_maps} and are discussed further in Sect. \ref{sec:mols_arise_shock}. 

The main peak of the emission is located toward the hot core, but these species also exhibit a clear secondary peak toward the southwest, at $\sim$5.1$''$ ($\sim$19000 au) from the center of the hot core. This emission is spatially coincident with that of the PN (2$-$1) transition observed by \cite{Fontani2024} (see their Fig. 1 and the green contours in our Fig. \ref{fig:mols_intro_shock_region}). Phosphorus-bearing molecules were identified in recent years as excellent tracers of shocks (e.g., \citealt{Lefloch2016, Mininni2018, Fontani2019, Rivilla2020a, Bergner2022, Lefloch2024}).
Moreover, this position is also coincident with bright emission of other well-known shock tracers, for example, SiO or sulfur-bearing species (\citealt{Fontani2024}), which trace the multiple molecular outflows that were identified in the region by \cite{Beltran2022a}. The distance to the hot core ensures that this position is free from the thermal processing of the protostars. This position, which is the G31.41 shock, therefore provides a suitable testbed for studying the chemical reservoir of freshly desorbed material from the icy grain mantles due to shocks.

\begin{figure*}
\centering
\includegraphics[scale=0.55]{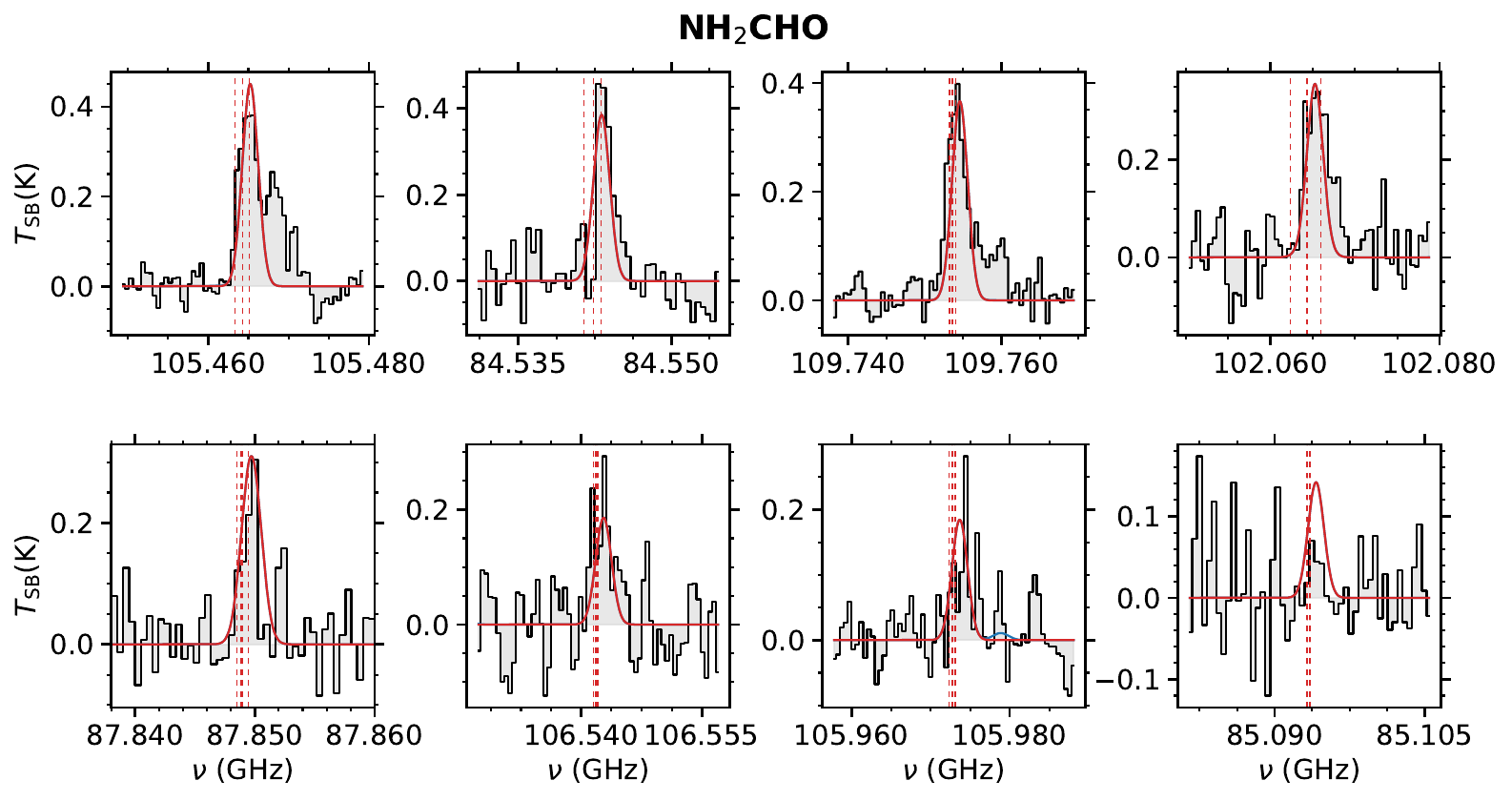}
\vspace{-2mm}
\caption{Transitions of NH$_2$CHO detected toward the G31.41 shock position. The black histogram and its gray shadow represent the observed spectrum. The red curve shows the best LTE fit of the individual species, and the blue curve shows the cumulative fit considering all detected species. The dashed red lines indicate the frequency of the transitions we fit. The plots are sorted by decreasing line intensity of the transitions.}
\label{fig:NH2CHO}
\end{figure*}

\begin{figure*}
\centering
\includegraphics[scale=0.55]{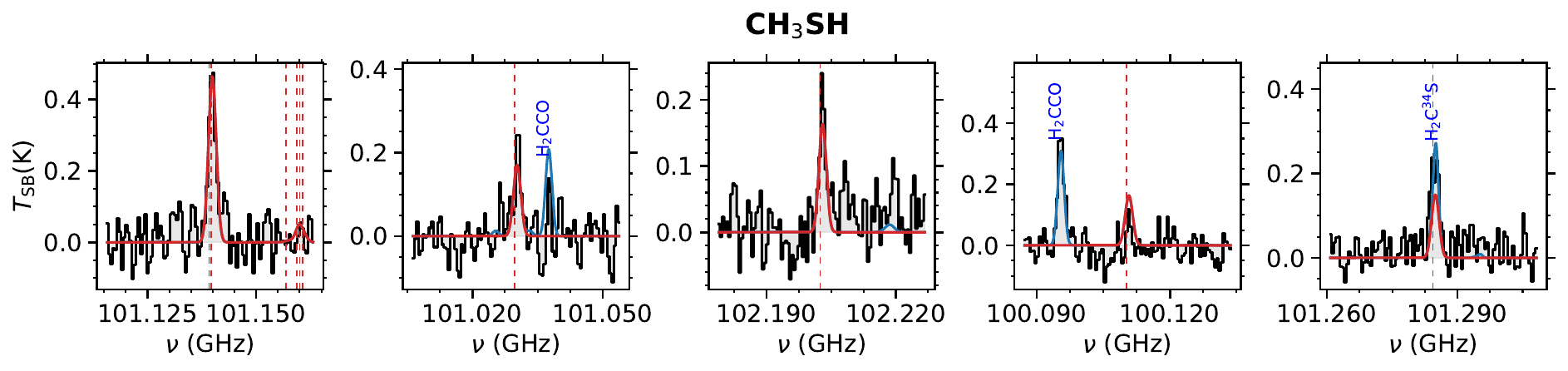}
\vspace{-2mm}
\caption{Transitions of CH$_3$SH detected toward the G31.41 shock position. The black histogram and its gray shadow represent the observed spectrum. The red curve shows the best LTE fit of the individual species, and the blue curve shows the cumulative fit considering all detected species. The dashed red lines indicate the frequency of the transitions we fit. The plots are sorted by decreasing line intensity of the transitions.}
\label{fig:CH3SH}
\end{figure*}

\subsection{Line identification and fitting}

\subsubsection{G31.41 shock position}
\label{sec:detceted_mols}

We extracted the spectra of the region delimited by the emission of the PN (2$-$1) transition observed by \cite{Fontani2024} (shown as a green polygon in Fig. \ref{fig:mols_intro_shock_region}). The resulting spectrum toward this position is significantly less crowded than that of the hot core. We therefore first extracted the spectra using the original cubes (with continuum), and we then subtracted the continuum with MADCUBA fitting baselines using line-free channels.

To perform the line identification and fitting, we used the Spectral Line Identification and Modeling (SLIM) tool within MADCUBA, which incorporates publicly available spectroscopic molecular catalogs: Cologne Database for Molecular Spectroscopy (CDMS\footnote{http://cdms.astro.uni-koeln.de/classic/}; \citealt{Muller2001, Muller2005, Endres2016}), Jet Propulsion Laboratory (JPL\footnote{https:// spec.jpl.nasa.gov/ftp/pub/catalog/ catdir.html}; \citealt{Pickett1998}), and Lille Spectroscopic Database (LSD\footnote{https://lsd.univ-lille.fr/}). SLIM assumes local thermodynamic equilibrium (LTE) to create a synthetic spectrum, taking the line opacity into account. SLIM compares the synthetic LTE spectrum with the observational spectrum and determines the best fit by applying the tool AUTOFIT, which is based on the Levenberg–Marquardt algorithm (see \citealt{Martin2019} for the description of the formalism). The physical parameters used in the fit are the molecular column density ($N$), the excitation temperature ($T_{\rm ex}$), the velocity ($\rm v$), the full width at half maximum (FWHM) of the line profile, and the source size. The latter was fixed to zero because we assumed that the emission fills the extracted  polygon.

After we identified the molecules, we selected the most unblended transitions to run AUTOFIT. When possible, we left the four parameters free. When the program did not converge when we left $T_{\rm ex}$ free, we fixed it to 20 K for some cases. This is an approximate average value obtained from the molecules for which convergence was reached (see Table \ref{tab:Poperties_molecules_on_G31_shock}), and it is also consistent with the value derived from the shock tracers analyzed by \cite{Fontani2024}.
For the fits for which we fixed the velocity, a value of 94 km s$^{-1}$ was used. This is the typical value derived from the fits for which the velocity was left as a free parameter (see Table \ref{tab:Poperties_molecules_on_G31_shock}). 
For the FWHM, we used the value of the CH$_3$CHO fit, 6.5 km s$^{-1}$, whose transitions are numerous, unblended, and intense (see Fig. \ref{fig:CH3CHO}). In some cases, a second component was required to obtain a better fit. We refer to the components of these molecules as broad or narrow depending on the FWHM of each component, which are listed in Table \ref{tab:Poperties_molecules_on_G31_shock}.

Because the most abundant molecules in the G31.41 shock might be optically thick (we assumed an arbitrary value of $\tau$ > 0.2), the retrieved values of the column density are lower limits. We therefore used the optically thinnest isotopolog whose fit was good to derive the column density of the main isotopolog by using the corresponding isotopic ratio. We used the same isotopic ratios as in Paper V \citep{Lopez-Gallifa2024}:
$\mathrm{^{12}C/^{13}C}$ = 39.5 $\pm$ 9.6 (\citealt{Milam2005, Yan2019}), $\mathrm{^{14}N/^{15}N}$ = 340 $\pm$ 90 (\citealt{Colzi2018}), $\mathrm{^{32}S/^{34}S}$ = 14.6 $\pm$ 2.1 (\citealt{Yu2020}), $\mathrm{^{16}O/^{18}O}$ = 330 $\pm$ 140 (\citealt{Wilson1999}), $\mathrm{^{34}S/^{33}S}$ = 5.9 $\pm$ 1.5 (\citealt{Yu2020}), $\mathrm{^{34}S/^{36}S}$ = 115 $\pm$ 17 (\citealt{Mauersberger1996}), and $\mathrm{^{18}O/^{17}O}$ = 3.6 $\pm$ 0.2 (\citealt{Wilson1999}).

When not enough unblended transitions were available to secure a detection, we derived upper limits to the column density. We used the brightest and unblended transition according to the LTE model and applied the upper limit tool of MADCUBA, which calculates the local rms of the spectrum ($\sigma$) by fitting a baseline and provides the column density for the upper limit of the integrated intensity at the 3$\sigma$ level (see more details in Sect. 3.4 of \citealt{Martin2019}). We fixed the other parameters of the upper limits following the aforementioned criteria (see Table \ref{tab:Poperties_molecules_on_G31_shock}).

We identified 30 new molecular species in the spectra and 18 isotopologs (see Table \ref{tab:Poperties_molecules_on_G31_shock} and \ref{tab:Fits_isotopologues_molecules_on_G31_shock}).
Therefore, the total number of molecules identified in the G31.41 shock is 35 molecules, considering the 5 additional species from \cite{Fontani2024} (SiO, SiS, SO, SO$_2$, PN, and PO). Moreover, we searched for another 20 species that were reported in the G31.41 core in previous GUAPOS works, but were not detected. In these cases, we derived the upper limits to the column density. The results for the other detected isotopologs are listed in Table \ref{tab:Fits_isotopologues_molecules_on_G31_shock}.

We describe the LTE fit of each detected molecule in detail in \href{https://zenodo.org/records/17613203?token=eyJhbGciOiJIUzUxMiJ9.eyJpZCI6IjAxOWRjMTY5LTNkZTYtNDhiZS04NzA2LTBiMWE1YzNiNmI0YSIsImRhdGEiOnt9LCJyYW5kb20iOiJjMjgzN2JkYTFjMDA0MDhmMDQyNTcwMTBhYmEzYzM0ZSJ9.enfIJaPU7GKZyCP8_Uhs1amywiqM5foSLAnZ6fooZahI8b6HFVqFovkdjdWFzcdic8wjooYvi0WP4Qw41KatcQ}{Appendix S 1}. When the molecule is optically thick, we present the fit of the isotopolog we used to derive the main isotopolog column density. The transitions we used to obtain each fit are listed in \href{https://zenodo.org/records/17613203?token=eyJhbGciOiJIUzUxMiJ9.eyJpZCI6IjAxOWRjMTY5LTNkZTYtNDhiZS04NzA2LTBiMWE1YzNiNmI0YSIsImRhdGEiOnt9LCJyYW5kb20iOiJjMjgzN2JkYTFjMDA0MDhmMDQyNTcwMTBhYmEzYzM0ZSJ9.enfIJaPU7GKZyCP8_Uhs1amywiqM5foSLAnZ6fooZahI8b6HFVqFovkdjdWFzcdic8wjooYvi0WP4Qw41KatcQ}{Table S 1} and are plotted in Figs. \ref{fig:CH3CHO}, \ref{fig:NH2CHO}, \ref{fig:CH3SH}, and \href{https://zenodo.org/records/17613203?token=eyJhbGciOiJIUzUxMiJ9.eyJpZCI6IjAxOWRjMTY5LTNkZTYtNDhiZS04NzA2LTBiMWE1YzNiNmI0YSIsImRhdGEiOnt9LCJyYW5kb20iOiJjMjgzN2JkYTFjMDA0MDhmMDQyNTcwMTBhYmEzYzM0ZSJ9.enfIJaPU7GKZyCP8_Uhs1amywiqM5foSLAnZ6fooZahI8b6HFVqFovkdjdWFzcdic8wjooYvi0WP4Qw41KatcQ}{Appendices S 3 and S 4}. The results of the fitting are summarized in 
Table \ref{tab:Poperties_molecules_on_G31_shock}. The spectroscopic information of all the molecules we used is summarized in \href{https://zenodo.org/records/17613203?token=eyJhbGciOiJIUzUxMiJ9.eyJpZCI6IjAxOWRjMTY5LTNkZTYtNDhiZS04NzA2LTBiMWE1YzNiNmI0YSIsImRhdGEiOnt9LCJyYW5kb20iOiJjMjgzN2JkYTFjMDA0MDhmMDQyNTcwMTBhYmEzYzM0ZSJ9.enfIJaPU7GKZyCP8_Uhs1amywiqM5foSLAnZ6fooZahI8b6HFVqFovkdjdWFzcdic8wjooYvi0WP4Qw41KatcQ}{Table S 3}.

\subsubsection{Additional molecules toward the G31.41 core}
\label{sec:detceted_mols_core}
To compare the chemical inventory of the G31.41 shock and that of the neighboring hot core, we also analyzed the emission of molecules toward G31.41 core that were identified in the G31.41 shock (Sect. \ref{sec:detceted_mols}) but not included in previous GUAPOS works. These species includes ten additional molecular species (and two additional isotopologs). We analyzed the spectra of the hot core following the same procedure as in Paper V (\citealt{Lopez-Gallifa2024}). When the AUTOFIT tool did not converge, we fixed the values of the velocity and/or the FWHM to 96.5 km s$^{-1}$ and 7.0 km s$^{-1}$, respectively, which are both different from the values used for the shock region. Similarly, for $T_{\rm ex}$, we used 50 and 150 K for molecules with < 6 atoms and $\geq$ 6 atoms, respectively. If the molecule was optically thick, we used the isotopologs
to obtain the corrected column density by applying the same isotopic ratios as in Paper V \citep{Lopez-Gallifa2024}. For the molecules that were not detected in the hot core, we computed upper limits to the column density using the brightest and most unblended transition according to the LTE model. Due to the chemical richness of the hot core, the spectrum barely contains line-free channels, and we were therefore unable to use the same MADCUBA procedure as toward the shock. In this case, as previously done in Paper V (\citealt{Lopez-Gallifa2024}), we fixed $T_{\rm ex}$, {\rm v}, and FWHM (to the values already explained in this section), and we increased the column density ($N$) until the LTE-predicted model was no longer visually compatible with the observed spectrum. 

We describe the LTE fits in detail in \href{https://zenodo.org/records/17613203?token=eyJhbGciOiJIUzUxMiJ9.eyJpZCI6IjAxOWRjMTY5LTNkZTYtNDhiZS04NzA2LTBiMWE1YzNiNmI0YSIsImRhdGEiOnt9LCJyYW5kb20iOiJjMjgzN2JkYTFjMDA0MDhmMDQyNTcwMTBhYmEzYzM0ZSJ9.enfIJaPU7GKZyCP8_Uhs1amywiqM5foSLAnZ6fooZahI8b6HFVqFovkdjdWFzcdic8wjooYvi0WP4Qw41KatcQ}{Appendix S 6}.
The transitions we used are listed in \href{https://zenodo.org/records/17613203?token=eyJhbGciOiJIUzUxMiJ9.eyJpZCI6IjAxOWRjMTY5LTNkZTYtNDhiZS04NzA2LTBiMWE1YzNiNmI0YSIsImRhdGEiOnt9LCJyYW5kb20iOiJjMjgzN2JkYTFjMDA0MDhmMDQyNTcwMTBhYmEzYzM0ZSJ9.enfIJaPU7GKZyCP8_Uhs1amywiqM5foSLAnZ6fooZahI8b6HFVqFovkdjdWFzcdic8wjooYvi0WP4Qw41KatcQ}{Table S 2}, and they are plotted in \href{https://zenodo.org/records/17613203?token=eyJhbGciOiJIUzUxMiJ9.eyJpZCI6IjAxOWRjMTY5LTNkZTYtNDhiZS04NzA2LTBiMWE1YzNiNmI0YSIsImRhdGEiOnt9LCJyYW5kb20iOiJjMjgzN2JkYTFjMDA0MDhmMDQyNTcwMTBhYmEzYzM0ZSJ9.enfIJaPU7GKZyCP8_Uhs1amywiqM5foSLAnZ6fooZahI8b6HFVqFovkdjdWFzcdic8wjooYvi0WP4Qw41KatcQ}{Appendices S 7 and S 8}. 
Table \ref{tab:Poperties_molecules_on_G31_core} summarizes the fit results.

\subsection{Deriving the molecular abundances toward the G31.41 shock}
\label{sec:H2_calc}

To derive the molecular abundances from the column densities, we obtained an estimate of the H$_2$ column density using CS.
This species is a good proxy of $N_{\rm H_2}$, as shown by \citet{Requena-Torres2006} in a survey of molecular clouds. CS is also a good reference for computing relative molecular ratios because its chemistry is nearly independent of the physical properties of the sources being barely enhanced in shocked gas (\citealt{Requena-Torres2006}) and UV photoresistant (\citet{Requena-Torres2008,Martin2008}.
Using the derived column density of CS in G31.41 shock and assuming the same molecular abundance found in the G31.41 core (7 $\times$ 10$^{-8}$ cm$^{-2}$ from Paper V, i.e., \citealt{Lopez-Gallifa2024}), we obtained the value $N$(H$_2$) = (1.2 $\pm$ 0.6) $\times$ 10$^{23}$ cm$^{-2}$. 
The abundances of the molecules were then obtained from the ratio $N/N_{\textrm{H}_2}$, and they are shown in Table \ref{tab:Poperties_molecules_on_G31_shock}. 

We also obtained an estimate of the average volume density of the G31.41 shock.
The area from which we extracted the spectra (depicted by the green contours in Fig. \ref{fig:mols_intro_shock_region}) is (5.4$''$)$^2$. The diameter of a circle with the same area is 2.6$''$. After deconvolving with the synthesized beam (1.2$''$) using the expression in Sect. 4.1 of \citet{Rivilla2017}, we obtained a deconvolved size of $\sim$ 2.3$''$, which translates into 0.042 pc or 8625 au 
(considering the heliocentric distance of 3.75 kpc obtained by \citealt{Immer2019}). Using this value and the $N$(H$_2$) derived previously, we obtain a volume density of $n$(H$_2$) of $\sim$2 $\times$ 10$^6$ cm$^{-3}$, which supports that the LTE conditions we assumed for the line analysis are indeed a good approximation.

\begin{figure}
\includegraphics[width=\columnwidth]{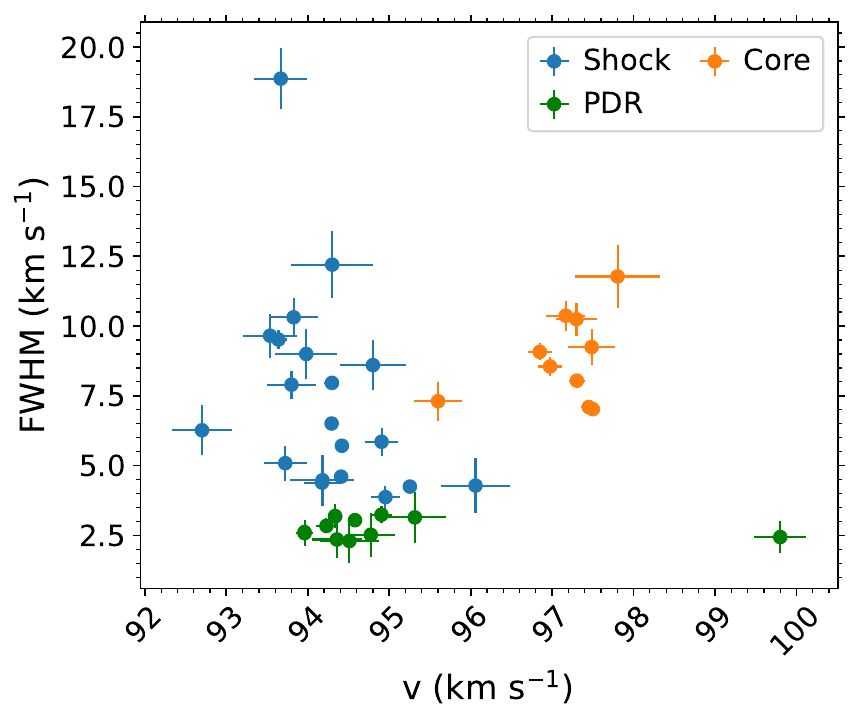}
\vspace{-2mm}
\caption{Comparison of the line widths and velocities of the molecules detected toward the G31.41 shock (from Table \ref{tab:Poperties_molecules_on_G31_shock} and from Paper IV) with those in the G31.41 core (from Table \ref{tab:Poperties_molecules_on_G31_core}; Papers I, II, III, and V). The species identified in the G31.41 shock are separated into two groups (see Sect. \ref{sec:mols_arise_shock} for details): those directly related with the shock (blue dots), and those likely related with an extended PDR in the line of sight (green dots). The molecules detected in the core are indicated with orange dots. We only plot the molecules for which FWHM and v  were left free in the fits.}
\label{fig:FWHM_v_shock_core}
\end{figure}

\begin{figure}
\includegraphics[width=\columnwidth]{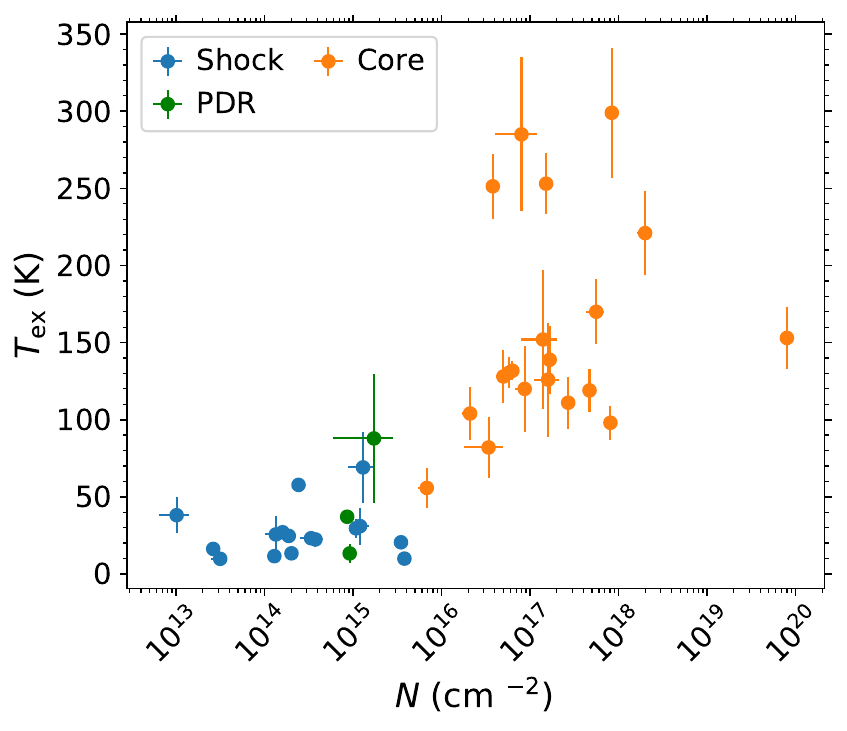}
\vspace{-2mm}
\caption{ Same as Fig. \ref{fig:FWHM_v_shock_core}, but for the excitation temperature ($T_{\mathrm{ex}}$) and column density ($N$). We only plot the molecules in which the parameters $T_{\mathrm{ex}}$ and $N$ converged.}
\label{fig:N_T_shock_core}
\end{figure}

\subsection{Origin of the detected molecules identified toward the G31.41 shock
}
\label{sec:mols_arise_shock}

In Sect. \ref{sec:detceted_mols} we identified the molecules in the spectra extracted toward the G31.41 shock position. 
Before the derived molecular abundances in the shock are compared with those in the hot core, it needs to be determined whether all these species are indeed physically related to the shock, or alternatively, if they arise from different gas that is located along line of sight. In order to distinguish between these two options, we first used the spatial distribution of the molecular emission shown in  Fig. \ref{fig:mols_intro_shock_region} and Appendix \ref{sec:Emission_maps}.
For N$_2$H$^+$ and HC$^{18}$O$^+$, unlike as shown in the maps, there is molecular emission from the hot core, but it is absorbed. 

Some of the species exhibit a clear peak in the G31.41 shock region: CH$_3$OH (narrow and broad component), CH$_3$CN, HNCO, CH$_3$OCH$_3$, NS, C$_2$H$_5$OH, and OCS (narrow and broad component). This suggests that they are physically related with the shocked gas. 
Other species show a different morphology, with emission that is more extended and elongated in the E-W direction southward of the G31.41 core. These are CCH, HNC, N$_2$H$^+$, HCO$^+$, c-C$_3$H$_2$, and CH$_3$CCH. 
The emission of these species resembles that of the gas that is traced by the 1.3 cm continuum (see Fig. 7 in \citealt{moscadelli2013}). This extended emission is likely related to a photodominated region (PDR) powered by the ultracompact HII region, which lies behind the G31.41 core on the line of sight and is located northeast of the G31.41 core (see Fig. \ref{fig:mols_intro_shock_region}). CCH and c-C$_3$H$_2$ are indeed typical tracers of PDRs (e.g., \citealt{Pety2005, Cuadrado2015}).
Therefore, these species might be more related to this extended gas component than to the shock. To confirm this, we compared the line widths (FWHM) of the species of these two groups. Figure \ref{fig:FWHM_v_shock_core} shows that the molecules that peak toward the shock span a FWHM range 4.2$^{+0.3}_{-0.3}$ - 18.9$^{+1.1}_{-1.1}$ km s$^{-1}$, while those showing a more extended emission are significantly narrower (green dots). They cover a FWHM range of 2.4$^{+0.6}_{-0.6}$ - 3.1$^{+0.3}_{-0.3}$ km s$^{-1}$, which supports a different origin of the molecules. 

We thus used the different FWHM ranges to classify the remaining molecules whose spatial distribution prevents a direct classification into each group. According to their derived FWHM, we therefore classify as molecules related to the shock HC$^{15}$N, H$_2$CNH, CH$_3$OH (both components), CH$_3$CN, NH$_2$CN, H$_2$CCO, HNCO, CH$_3$CHO, C$^{36}$S, NH$_2$CHO, HCS$^+$, H$_2$CS (broad component), CH$_3$OCH$_3$, NS, C$_2$H$_5$OH, CH$_3$SH, H$^{13}$CCCN, HC$^{13}$CCN, C$_2$H$_3$CN,  CH$_3$OCHO, OCS, and HC$_5$N. On the other hand, NH$_2$D (both components), CCH, HN$^{13}$C, $^{13}$C$^{18}$O, N$_2$H$^+$, HC$^{18}$O$^+$, H$_2$CO, c-C$_3$H$_2$, CH$_3$CCN, and H$_2$CS (narrow component) appear to be more related with the extended photodominated gas. 
In our discussion of the G31.41 shock, we only consider the molecules classified as associated with the shocked gas.

We also compared the physical properties of the molecular emission toward the G31.41 shock (Sect. \ref{sec:detceted_mols}) with those of the hot core. We obtained the physical properties in G31.41 core from previous GUAPOS works and Sect. \ref{sec:detceted_mols_core}).

As shown in Fig. \ref{fig:FWHM_v_shock_core} (FWHM vs. v), the velocity of the molecules in the shock is blueshifted ({\rm v}$\simeq$92$-$95.5 km s$^{-1}$, except for the narrow component of NH$_2$D) with respect to that of the hot core ({\rm v}$\simeq$95.5$-$98 km s$^{-1}$). This suggests that they are associated with the blueshifted lobe of one of the SiO outflows observed in the region (see Fig. 1 in \citealt{Fontani2024}, \citealt{Beltran2018, Beltran2022a}).

Figure \ref{fig:N_T_shock_core} ($T_{\rm ex}$ vs. $N$) shows that the excitation temperatures derived in the shock are also significantly lower (10$-$70 K) than those in the hot core (70$-$300 K). The former are similar to those reported in the low-mass protostellar outflow L1157-B1 (10$-$80 K, \citealt{Lefloch2017}). Since the average volume density of the shock and core in G31.41 are high ($\sim$2 $\times$ 10$^6$ cm$^{-3}$; see Sect. \ref{sec:H2_calc}; and $\sim$10$^8$ cm$^{-3}$; \citealt{Mininni2020}, respectively), the lines are expected to thermalize, and the $T_{\rm ex}$ range in the two sources is therefore expected to reflect the kinetic temperature of the gas. This confirms that the gas in the shock is colder, which suggests that it is already in a post-shock phase in which the temperature has decreased, and it is not affected by the heating from the protostars.

The column densities in the shock position are about 10$^{13}$ - 10$^{15}$ cm$^{-2}$. This is lower by some orders of magnitude than those found in the hot core (10$^{16}$ - 10$^{17}$ cm$^{-2}$).

\begin{figure*}
\includegraphics[width=0.9\textwidth]{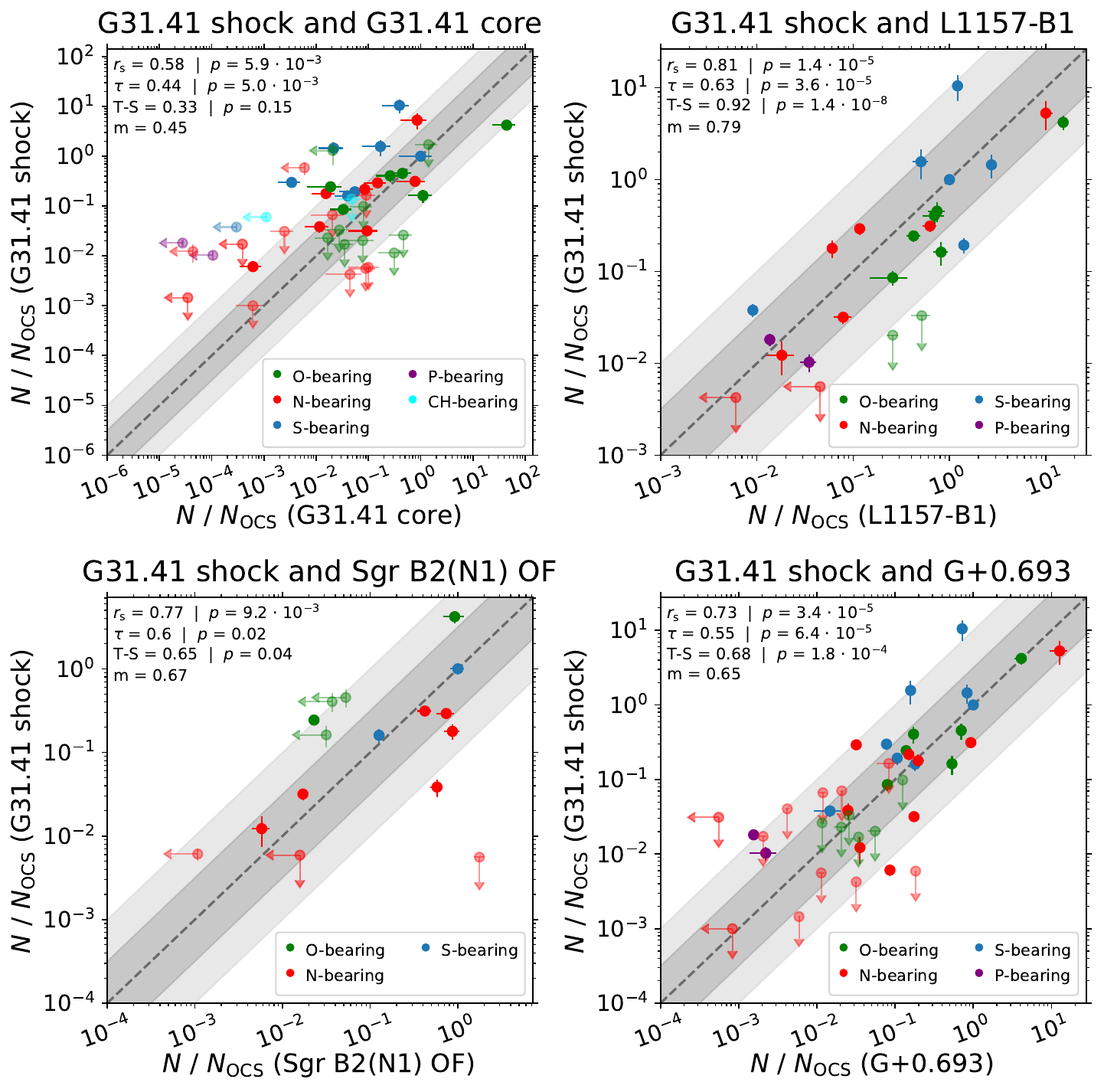}
\vspace{-2mm}
\caption{
{Comparison of the molecular ratios with respect to \ch{OCS} in the G31.41 shock with those in the G31.41 core (\textit{upper left}), L1157-B1 shock (\textit{upper right}), Sgr B2(N1) OF shock (\textit{lower left}), and in the G+0.693 molecular cloud (\textit{lower right}). The different families of molecules (O-, N- S-, P-, and CH-bearing) are denoted with different colors. The upper limits for undetected species are indicated with arrows. The light and dark shaded gray area corresponds to half and one order of magnitude scatter from the y=x relation denoted by the dashed line. The results of the correlation tests are indicated in the upper left corner of each panel: $r_{\mathrm{s}}$ (Spearman), $\tau$ (Kendall), and T-S (Theil-Sen), with their associated $p$-values, and the average $m$ (for details, see Sect. \ref{sec:Comparison of molecular abundances}).}
}
\label{fig:Correlaciones_fuentes_colores}
\end{figure*}

\section{G31.41 shock versus hot core, and comparison with other sources}
\label{sec:comparison}

The chemical census of the G31.41 shock obtained from this work and from \citet{Fontani2024} allowed us to directly compare the derived abundances with those of the G31.41 core. Moreover, we also compared them with three additional interstellar sources whose chemistry is thought to be dominated by shocks: the protostellar outflows L1157-B1 and Sgr B2(N1) OF powered by a low-mass and a high-mass protostar, respectively, and the molecular cloud G+0.693-0.027 (hereafter G+0.693). These three sources were previously studied in detail, and their molecular inventory includes tens of molecules whose  column densities were derived in an homogeneous way. Hence, this allowed us to perform a complete comparative study.

L1157-B1 is a shock created by the  blueshifted southern lobe of the outflow driven by the class 0 protostar L1157-mm \citep{Bachiller2001}.  
 We collected the column densities from single-dish observations (IRAM 30m or Herschel) presented by
\cite{Arce2008, Sugimura2011, Lefloch2012, Mendoza2014, Podio2014, Gomez-Ruiz2015, Lefloch2016, Lefloch2017, Mendoza2018, Holdship2019, Lefloch2021}. 
All these works performed a similar analysis, assuming a similar source size for the molecular emission of 18$''$or 20$''$.
For CH$_3$CN, we analyzed the publicly available data from the Astrochemical Surveys At IRAM (ASAI) project\footnote{https://www.oan.es/asai/} with MADCUBA. We converted the temperature scale of the data from the antenna temperature ($T_A*$) into the main-beam temperature ($T_{\rm mb}$), and we assumed a source size of 18$''$ for consistency. The resulting column density is reported in Table \ref{tab:Molecules_used_each_source}.
We considered a total of 22 detected molecules and two upper limits (see Table \ref{tab:Molecules_used_each_source}).

Sgr B2(N1) OF is a shock-affected region that is thought to have been produced by the encounter of the blueshifted (and most intense) lobe of a molecular outflow with the gas of the parental hot core in the N1SE1 region of Sgr B2 (see Fig. 1 of \citealt{Busch2024}). 
We obtained the column densities of 11 detected molecules (and an additional five upper limits) from the analysis by \cite{Busch2024} (see our Table \ref{tab:Molecules_used_each_source}).

G+0.693 is a molecular cloud that is also located in the Sgr B2 region of the Galaxy center, which is thought to be affected by large-scale shocks from cloud-cloud collisions (\citealt{Zeng2020, Colzi2024}). We obtained the values for the column densities from many works that analyzed an ultradeep spectral survey toward this cloud in detail (e.g., \citealt{Zeng2018, Rodriguez-almeida2021a,Rivilla2023}).
We considered 40 detected molecules and two upper limits (see Table \ref{tab:Molecules_used_each_source}).

For the comparative study, we considered a total of 45 molecular species, listed in Table \ref{tab:Molecules_used_each_source}, which also includes the information about the detection toward the other shock-dominated sources and the derived column densities. We obtained the column densities of the G31.41 core from Papers I, II, III, and V (\citealt{Mininni2020, Colzi2021, Mininni2023, Lopez-Gallifa2024}, respectively) and from this work (see Table \ref{tab:Poperties_molecules_on_G31_core}).

\subsection{Comparison of the molecular ratios}
\label{sec:Comparison of molecular abundances}

For the comparison between the sources, we normalized the column densities with those of the narrow component of OCS (Table \ref{tab:Molecules_used_each_source}). We chose this molecule because it presents a clear enhancement toward the G31.41 shock, indicating a clear physical relation (see Fig. \ref{fig:mols_intro_shock_region}), and it is detected in the five sources we considered. 

Figure \ref{fig:Correlaciones_fuentes_colores} compares the molecular ratios in the G31.41 shock with those found in the other four sources. For the two molecules related to the G31.41 shock that were fit with two components (CH$_3$OH and OCS), we considered the component whose velocity and line width values were more similar to those of the other molecules in the shock (Fig. \ref{fig:FWHM_v_shock_core}). We note in any case that the column density of the two components is similar (Table \ref{tab:Poperties_molecules_on_G31_shock}), and this choice therefore does not affect the results.

To allow a quantitative interpretation of the possible correlations, we performed three different statistical tests: Spearman (r$_\textrm{s}$), Kendall ($\tau$), and Theil-Sen (T-S). The Spearman and Kendall methods quantify whether the points increase monotonically, while the Theil-Sen test determines how well the points can be fit linearly. The results of the correlation tests range from 1 (perfect correlation) to -1 (anticorrelation), with 0 meaning no correlation (for more details, see Appendix E of \citealt{Lopez-Gallifa2024}). 
These tests were applied to the points that correspond to the detected molecules in both sources of each pair. To verify the reliability of the tests, we also calculated their associated $p$-value, which quantifies 
the goodness of the correlation. The $p$-value varied from 0 (perfectly reliable correlation coefficient value) to 1 (unreliable value). 
The results of each test were then combined to simplify its interpretation. We defined the average of the statistical tests as

\begin{equation}
    m = \frac{r_{\mathrm{s}} + \tau + \textrm{T-S}}{3} .
\end{equation}

In our further discussion, we refer to $m$ to interpret the results arising from the correlations, using the following criteria:
$m\geq$ 0.9 (excellent correlation), 0.9 $>m\geq$ 0.7 (very good correlation), 0.7 $>m\geq$ 0.5 (good correlation), 0.5 $>m\geq$ 0.3 (poor correlation), and 0.3 $>m$ (no correlation). In the following sections, we describe the results of the comparative analysis considering the whole sample of molecules, and also separating them into O-, N- and S-/P-bearing families because we have enough molecules to analyze them separately.

\subsubsection{Full chemical reservoir}
\label{sec:full_chemical_reservoir}

We first discuss the results we found considering all the molecules (see Fig \ref{fig:Correlaciones_fuentes_colores}).
Interestingly, the worst correlation was found between the two sources located in the same region: the G31.41 shock and the G31.41 core. The average correlation coefficient is $m$=0.45, based on 21 molecules, which indicates a poor correlation. The molecular ratios of the G31.41 shock correlate significantly better with that of the other shock-dominated regions: $m$=0.79 (20 molecules) with L1157-B1, $m$=0.67 with Sgr B2(N1) OF (10 molecules), and $m$=0.65 (25 molecules) in G+0.693.

\subsubsection{O-bearing species}
\label{subsec:O-bearing species}

Figure \ref{fig:Correlaciones_fuentes_O} shows the molecular ratios considering O-bearing species alone, normalized with \ch{CH3OH}. We found similar results in general to those considering all the species. 
The correlation between the G31.41 shock and the G31.41 core is poor ($m$=0.48 with $p$ values < 0.38, based on six molecules). Moreover, the ratios of the detected molecules in the shock are always higher than those in the hot core. 
The best correlations are found between the
G31.41 shock and G+0.693, with $m$=0.84 (six molecules), and between the G31.41 shock and L1157-B1 ($m$=0.63, six molecules).
We note that the $p$ values are higher than those obtained for the full chemical reservoir (see Sect. \ref{sec:full_chemical_reservoir}), probably because the number of molecules for the comparison is smaller.
We did not perform the correlation test between the G31.41 shock and Sgr B2(N1) OF because only two molecules are available.
The molecular ratios in the G31.41 shock compared to those in the other three shocked regions are always similar within half an order of magnitude (dark gray shaded area in Fig. \ref{fig:Correlaciones_fuentes_O}).

\subsubsection{N-bearing species}
\label{subsec:N-bearing species}

Figure \ref{fig:Correlaciones_fuentes_N} shows the molecular ratios of N-bearing species, normalized with \ch{CH3CN}. In this case, compared to the O-bearing family, we found different results. For this family, the correlation between the G31.41 shock and the hot core is very good, with $m$=0.83 (based on eight molecules). Unlike for the O-bearing species, the ratios of N-bearing species are similar between these sources within one order of magnitude (shaded area in Fig. \ref{fig:Correlaciones_fuentes_N}).
The correlation between the G31.41 shock and L1157-B1 is excellent ($m$=0.93, six molecules), while the correlations with Sgr B2(N1) OF and G+0.693 are worse ($m$=0.53 and 0.51, based on six and nine molecules, respectively). In particular, the ratios in the G31.41 shock are clearly significantly lower than in G+0.693.

\subsubsection{S- and P-bearing species}
\label{subsec:S-bearing species}

Figure \ref{fig:Correlaciones_fuentes_S} shows the molecular ratios of S- and P-bearing species, normalized with OCS. 
The best correlation is found between the G31.41 shock and G+0.693 ($m$=0.66, based on ten molecules). The ratios of the species are slightly higher in the G31.41 shock within the gray shaded area (with the only exception of CS, which is only slightly higher).

The correlations with the G31.41 core and L1157-B1 are poor ($m$=0.31 and 0.43, based on seven and eight molecules, respectively). 
Moreover, we note that these correlations have high values of $p$, and hence, the results are not reliable. In the latter case, the ratios are within one order of magnitude (gray shaded area in the upper right panel in  Fig. \ref{fig:Correlaciones_fuentes_S}). In the former case, the ratios in the G31.41 shock are always higher than those in the core, similar to the O-bearing molecules, and their dispersion is higher. We did not perform the correlation test between the G31.41 shock and Sgr B2(N1) OF because only two molecules are available.

\section{Discussion}
\label{sec:discuss}

\subsection{Chemistry in shock regions versus hot cores}
\label{sec:Chemistry in shock regions vs. hot cores}

The spectral survey provided by the GUAPOS project has allowed us to study the molecular inventory of two regions in detail that share the same natal environment in the G31.41 high-mass star-forming region, but were subject to different physical events and feedback that have imprinted their chemistry: the hot core, and a shock-dominated position. 
In the hot core, the heating produced by the embedded protostars, which can reach hundreds of K (see orange dots in Fig. \ref{fig:N_T_shock_core}), is expected to be continuous and smooth, lasting for relatively long timescales of 10$^{4} - $10$^{6}$ yr (see, e.g., the warm-up models in \citealt{Garrod2013a}). This thermal processing can affect the chemistry in several ways: (i) It increases the mobility of radicals on dust grains, facilitating further chemical reactions. (ii) It desorbs the molecules formed on the surfaces into the gas. (iii) It might allow hot gas-phase chemistry. All these effects can alter the original chemical composition (and hence abundance ratios) on the icy mantles. Chemical models mimicking the evolution of a hot core (e.g., \citealt{Garrod2013a,Garrod2022}) indeed showed that the abundances of the species can vary by several orders of magnitude in the grains before they thermally desorb, and they also vary in the hot gas-phase after desorption.

In the case of shocks, although the shock front is able to increase to even higher temperatures compared with those of hot cores, the heating is much shorter in time. As shown in the physical model by \citet{Burkhardt2019}, the timescales for grain heating, sputtering from the grains, and the hot-gas phase last some hundreds of yr, a few 10$^{2}$ yr and 10$^{3}$ yr, respectively.
These short timescales - compared to those of warming in hot cores - might limit the chemistry that can take place on the grain surface and in the gas phase. The chemical models of shocks in molecular outflows by \citet{Burkhardt2019} indeed showed that for several molecules (e.g., CO, HCN, \ch{CH3OH}, \ch{H2CO}, \ch{HCOOH}, and \ch{CH3OCH3}, all of them included in our comparative study), there is little post-shock chemistry. Consequently, the derived gas-phase abundances are a good proxy (at least as a first approximation) of those in the grain mantles. The authors also noted, however, that some species (\ch{NH2CHO}, \ch{CH3OCHO}, or \ch{CH3CHO}) exhibited significant abundance enhancements during the post-shock gas-phase. We discuss this possibility in Sect. \ref{sec:Universality of shock-dominated chemistry}.

The molecular inventories of the hot core and the shock are poorly correlated (Sect. \ref{sec:full_chemical_reservoir}), with the exception of N-bearing molecules. Since it is expected that the initial chemical composition of the icy grain mantles of the whole natal environment is the same, this is an observational confirmation that the two regions were subject to different chemical processing.

N-bearing species show a good correlation between the G31.41 shock and the hot core. 
A similar result was found by \citet{Busch2024} from analyzing the shocked outflows in Sgr B2(N1), in which the derived abundances of N-bearing molecules were similar to those in the associated hot core. The physical evolution of hot cores and shocked regions is very different, and this might therefore indicate a preferential formation of N-bearing molecules at earlier evolutionary stages than the other families. This has been proposed, for instance, for NH$_2$CN in the hot-core models presented by \citet{Garrod2022}, in which the net production of this species occurs almost entirely at very early times in the model, when the protostar has not yet started to heat the environment.

\subsection{Universality of shock-dominated chemistry}
\label{sec:Universality of shock-dominated chemistry}

Interestingly, the molecular ratios of the G31.41 shock correlate significantly better with other shock-dominated regions than with the G31.41 core. This similarity is striking because the shocks we considered are significantly different in terms of powering source, environment, and size. 
The G31.41 shock, Sgr B2(N1) OF, and  L1157-B1  are shocks of protostellar origin, namely related with molecular outflows powered by high-mass protostars (the two former) and a low-mass protostar (the latter). 
The shocks in G+0.693 are thought to be the result of cloud-cloud collisions.
G31.41 and L1157-B1 are located in different regions of the Galactic disk, while Sgr B2(N1) OF and G+0.693 are located in the Galactic center.
Moreover, they all have very different sizes: $\sim$0.02-0.04 pc (L1157-B1 and G31.41 shock), $\sim$0.1 pc (Sgr B2(N1) OF), and $\sim$1 pc (G+0.693). Despite all these differences, however, the molecular ratios of these four sources correlate well.

The fact that similar molecular ratios are present in these different shocked regions, which are unlikely to be at the same post-shock age, indicates that the gas-phase chemistry after ejection from grains is negligible. 
As an example, no significant differences (within one order of magnitude) are found in the ratios of \ch{CH3OCHO} and \ch{CH3CHO} for the four shocked regions (see the upper right and lower panels of Fig. \ref{fig:Correlaciones_fuentes_O}), in contrast to the larger variations that are expected if relevant gas-phase chemistry were in place, as shown by the models by \citet{Burkhardt2019}.
This might indicate that the molecular content has been freshly desorbed, and that there has not been enough time to develop a significant gas-phase chemistry. If this is the case, the gas-phase material we detected represents a good proxy of the ice-mantle composition of the different regions. Since the molecular ratios do not vary strongly, this would accordingly imply that the mantle reservoir is similar, regardless of the Galactic environment. 
\citet{Requena-Torres2006} also reached this conclusion by comparing the abundances of a handful of O-bearing COMs toward tens of molecular clouds in the Galactic center with Galactic disk hot cores.

\section{Summary and conclusions}
\label{sec:conclusions}

We have provided an outlook on the chemical inventory of a shock region in the high-mass star-forming region G31.41+0.31, where we derived the column density of 30 species (plus 18 isotopologs) using data from the GUAPOS project. 
In total, we used 35 molecules (adding 5 detected molecules from Paper IV) to perform the comparative study. 

Using the large number of detected molecules, we compared the molecular inventory of the shock region of G31.41 with the hot-core region of the same source (in which we complemented previous works by analyzing six additional molecules). This comparison allowed us to investigate the differences between the desorbed molecular content due to the heating of stars (G31.41 core) and due to a shock created by an outflow (G31.41 shock), while having the same initial conditions on the grains, that is, the same natal environment. The results show that the correlation between the two regions is poor, except for N-bearing molecules. This suggests that the initial grain-mantle composition is processed differently in the hot core compared to the shock region. This divergence might be due to the different physical conditions of the two environments. A high-mass star-forming region is expected to warm the grain surface steadily and during longer timescales than shock regions. In addition, the temperature change in a shocked region is more abrupt than in hot cores. The only species that do not follow this trend are N-bearing species, and this might indicate that they are formed in earlier evolutionary stages.

We also compared the G31.41 shock region with another three shock-dominated regions: a low-mass protostellar outflow (L1157-B1), a high-mass protostellar outflow (Sgr B2(N1) OF), and a Galactic center shock-dominated molecular cloud (G+0.693). The results show a better correlation with other shock-dominated regions compare to the hot core. Despite the different sizes, environments, and Galactic locations, this good correlation might indicate that the molecular composition of the grains barely changes after they are desorbed, which would imply a negligible gas-phase chemistry. This suggests that the molecular content of shock-dominated regions can be a good proxy of the composition of the grain surfaces. Taking into account that there are no significant variations in the correlations between the different shock-dominated regions, we suggest that the molecular ratios in the mantle reservoir of the grains are similar, regardless of the Galactic environment.

\section*{Data availability}

The data presented in this work are available upon reasonable request. Appendix S is in the supplementary information available on  \href{https://zenodo.org/records/17613203?token=eyJhbGciOiJIUzUxMiJ9.eyJpZCI6IjAxOWRjMTY5LTNkZTYtNDhiZS04NzA2LTBiMWE1YzNiNmI0YSIsImRhdGEiOnt9LCJyYW5kb20iOiJjMjgzN2JkYTFjMDA0MDhmMDQyNTcwMTBhYmEzYzM0ZSJ9.enfIJaPU7GKZyCP8_Uhs1amywiqM5foSLAnZ6fooZahI8b6HFVqFovkdjdWFzcdic8wjooYvi0WP4Qw41KatcQ}{Zenodo}.

\begin{acknowledgements}
The table of molecular transitions were made thanks to the Python library richvalues created by Andrés Megías. This paper makes use of the following ALMA data: ADS/JAO.ALMA$\#$2017.1.00501.S. ALMA is a partnership of ESO (representing its member states), NSF (USA) and NINS (Japan), together with NRC (Canada), MOST and ASIAA (Taiwan), and KASI (Republic of Korea), in cooperation with the Republic of Chile. The Joint ALMA Observatory is operated by ESO, AUI/NRAO and NAOJ.

A. L-G., V.M.R. and L.C. acknowledges support from the grant PID2022-136814NB-I00 by the Spanish Ministry of Science, Innovation and Universities/State Agency of Research MICIU/AEI/10.13039/501100011033 and by ERDF, UE; and from the Consejo Superior de Investigaciones Cient{\'i}ficas (CSIC) and the Centro de Astrobiolog{\'i}a (CAB) through the project 20225AT015 (Proyectos intramurales especiales del CSIC). V.M.R. also acknowledges support from the grant RYC2020-029387-I funded by MICIU/AEI/10.13039/501100011033 and by "ESF, Investing in your future" and from the grant CNS2023-144464 funded by MICIU/AEI/10.13039/501100011033 and by “European Union NextGenerationEU/PRTR”. A.S.-M.\ acknowledges support from the RyC2021-032892-I grant funded by MCIN/AEI/10.13039/501100011033 and by the European Union `Next GenerationEU’/PRTR, as well as the program Unidad de Excelencia Mar\'ia de Maeztu CEX2020-001058-M, and support from the PID2020-117710GB-I00 (MCI-AEI-FEDER, UE).

\end{acknowledgements}

\bibliographystyle{aa}
\bibliography{bibliography}

\appendix

\onecolumn

\section{Emission maps of the detected molecules in G31.41 shock}
\label{sec:Emission_maps}

The spatial distribution of the molecular emission of the detected molecules toward G31.41 star-forming region (see Table \ref{tab:Poperties_molecules_on_G31_shock}), that have not been presented in the main text, is shown in Fig. \ref{fig:mols_shock_region}.

\begin{figure*}[ht]

\includegraphics[scale=0.39]{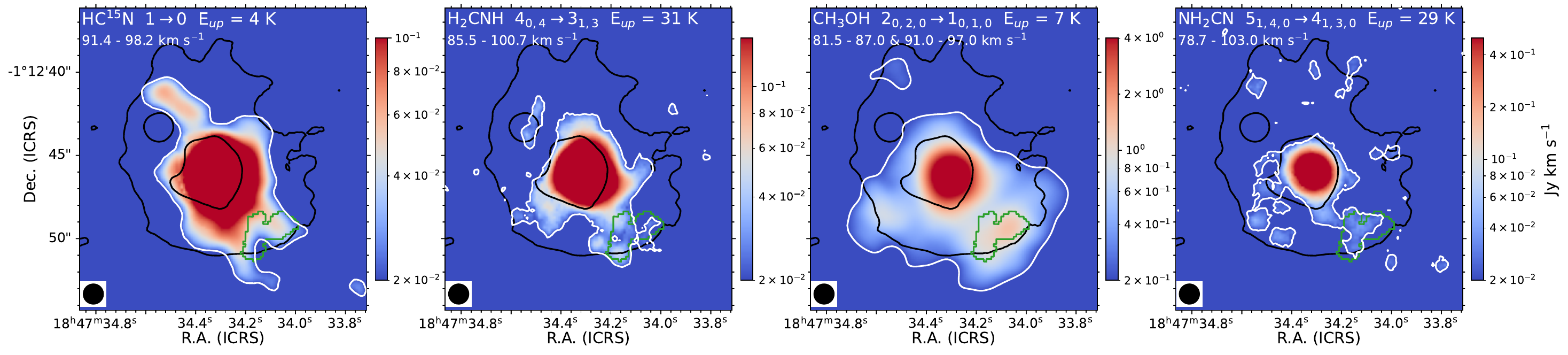}
\includegraphics[scale=0.39]{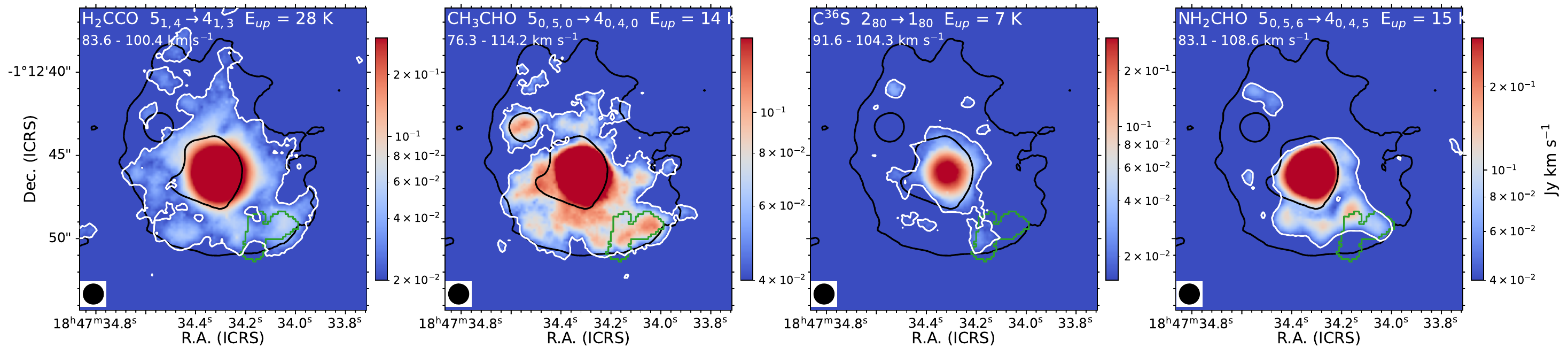}
\includegraphics[scale=0.39]{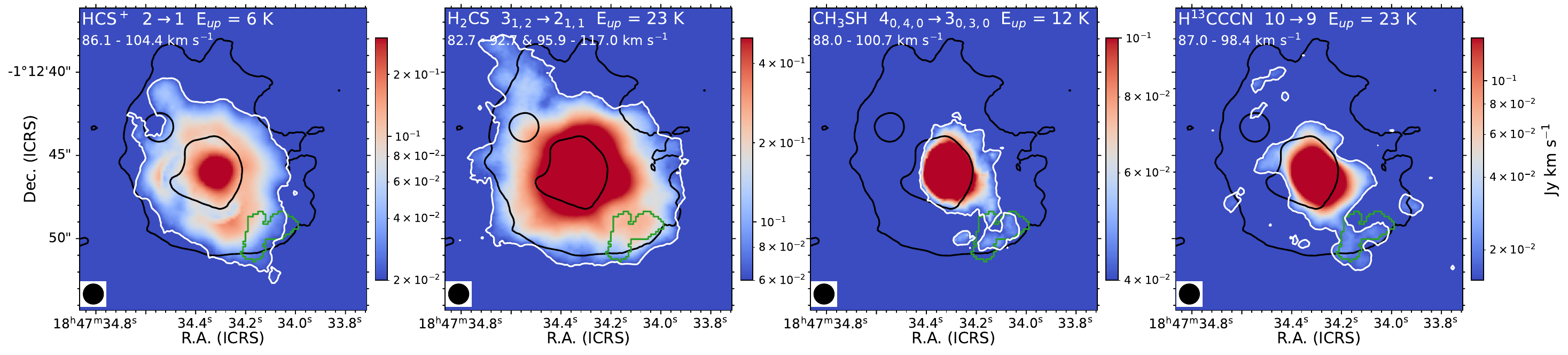}
\includegraphics[scale=0.39]{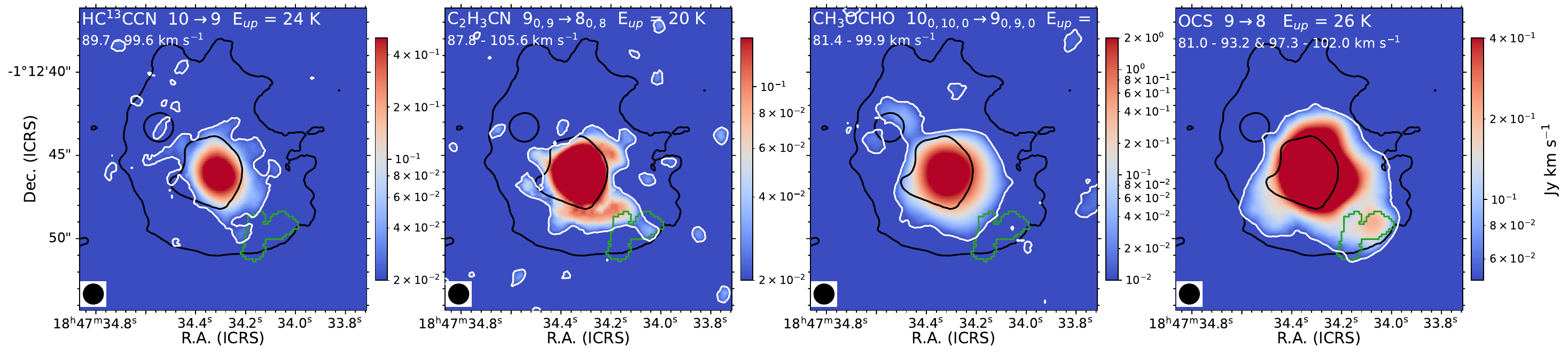}

\caption{Emission maps of detected molecules towards G31.41 shock. On each panel in the upper part of the panels is shown in white the molecule name, the molecular transition, the energy of the upper level and the velocity range. In addition, in green is labeled the G31.41 shock region according to region 2 of Fig. 1 in \cite{Fontani2024}. The ultracompact HII region and the hot core of G31.41 are also labeled. The 3 and 20 $\sigma$ contour levels of the 3mm continuum map at 98.5 GHz from \cite{Mininni2020} are plotted in black. The minimum flux level of the color bar of each panel is indicated by the white contours.}
\label{fig:mols_shock_region}
\end{figure*}

\begin{figure*} \ContinuedFloat
\includegraphics[scale=0.39]{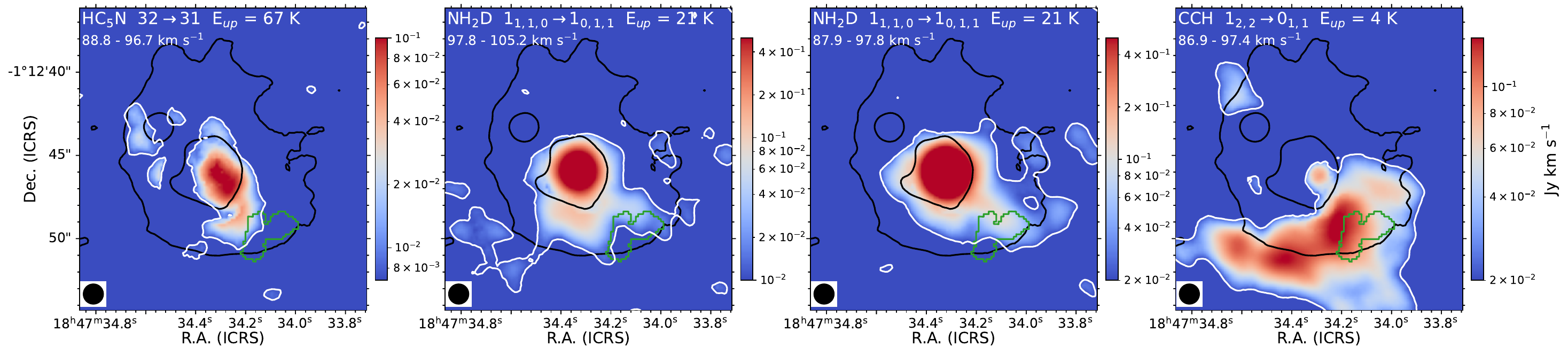}
\includegraphics[scale=0.39]{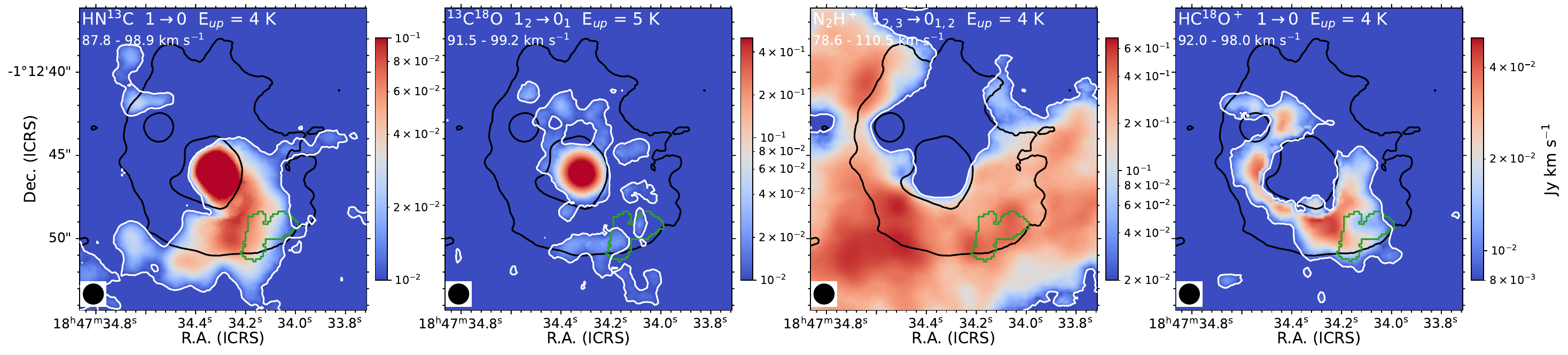}
\includegraphics[scale=0.39]{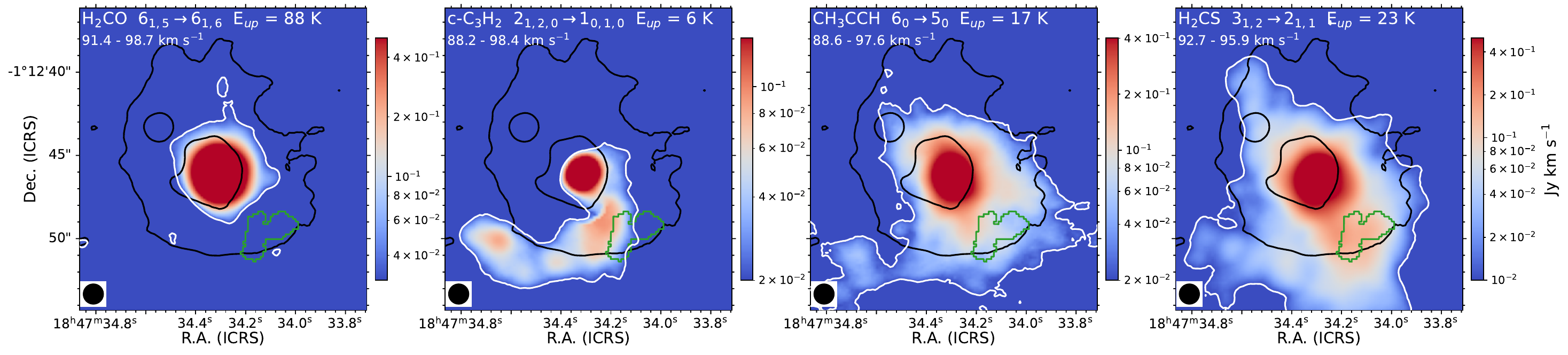}
\caption{Continued.}
\end{figure*}

\section{SLIM fits}

Table \ref{tab:Poperties_molecules_on_G31_shock} and \ref{tab:Poperties_molecules_on_G31_core} list the SLIM fits of the isotopologs analyzed toward G31.41 shock and core, respectively. Table \ref{tab:Fits_isotopologues_molecules_on_G31_shock} lists the SLIM fits of the isotopologs analyzed towards G31.41 shock that were not used to derive the abundances of the main isotopologs. The three tables are ordered by increasing molecular mass.

\setlength{\tabcolsep}{3.5pt}

\begin{longtable}{cccccccc}

\caption{\label{tab:Poperties_molecules_on_G31_shock} Results of the MADCUBA-SLIM fits of the molecules analyzed toward the G31.41 shock, ordered by increasing molecular mass.  
}
\\
\hline
        \hline
     Formula &  $\textit{T}_{\textrm{ex}}$ & \textit{N} & $\rm v$ & FWHM & $\tau_{max}^a$ & $N/N^b_{\textrm{H}_2}$ & Figure \\ 
     & (K) & ($\times$10$^{13}$ cm$^{-2}$) & (km s$^{-1})$ & (km s$^{-1})$ & & ($\times$10$^{-10}$ cm$^{-2}$) & \\ \hline \endfirsthead
     \caption{Continued.}\\
\hline
     Formula &  $\textit{T}_{\textrm{ex}}$ & \textit{N} & $\rm v$ & FWHM & $\tau_{max}^a$ & $N/N^b_{\textrm{H}_2}$ & Figure \\ 
     & (K) & ($\times$10$^{13}$ cm$^{-2}$) & (km s$^{-1})$ & (km s$^{-1})$ & & ($\times$10$^{-10}$ cm$^{-2}$) & \\ \hline 
   \endhead
    \endfoot
    
         \hline \noalign{\smallskip}
    \multicolumn{8}{p{17cm}}{\textbf{Notes.} The table is separated into three parts to distinguish molecules: (i) detected and related with shocked gas; (ii) detected in the shocked region but likely associated with extended photo-dominated gas (PDR); and (iii) not detected in the shocked region, for which column density upper limits are provided (marked with <; see Sect. \ref{sec:mols_arise_shock} for details). For optically thick molecules, we include the optically thin isotopolog that we have used to derive the column density of the main isotopolog (the results of the other isotopologs are presented in Table \ref{tab:Fits_isotopologues_molecules_on_G31_shock} instead). The resulting physical parameters, along with their associated uncertainties, are presented. The values of the parameters that were kept fixed are shown without uncertainties.  
We also present the maximum line opacity of the transitions of each molecule ($\tau_{max}$), and the molecular abundances compared to H$_2$. The transitions of each species used to perform the fits are listed in \href{https://zenodo.org/records/17613203?token=eyJhbGciOiJIUzUxMiJ9.eyJpZCI6IjAxOWRjMTY5LTNkZTYtNDhiZS04NzA2LTBiMWE1YzNiNmI0YSIsImRhdGEiOnt9LCJyYW5kb20iOiJjMjgzN2JkYTFjMDA0MDhmMDQyNTcwMTBhYmEzYzM0ZSJ9.enfIJaPU7GKZyCP8_Uhs1amywiqM5foSLAnZ6fooZahI8b6HFVqFovkdjdWFzcdic8wjooYvi0WP4Qw41KatcQ}{Table S 1}. The isotopic ratios and the $N$(H$_2$) used are discussed in Sect. \ref{sec:data analysis}. $^a$ Maximum $\tau$ of all the transitions detected for each molecule. $^b$ $N$(H$_2$) = (1.2 $\pm$ 0.6) $\times$ 10$^{23}$ cm$^{-2}$ of the G31.41 shock, more details in Sect. \ref{sec:H2_calc}. $^c$ Following the criteria based on the FWHM to distinguish if the molecule arises from the shock or not, we have added the two line components of H$_2$CS in different parts of the table. $^d$ The SLIM fit of the molecule has two components.}
     \endlastfoot

\multicolumn{8}{c}{Detected molecules related with the shock}
\tabularnewline \hline  
HCN & & $440 \pm 140$ & &  &  & $(4 \pm 2) \times 10^2$ & \href{https://zenodo.org/records/17613203?token=eyJhbGciOiJIUzUxMiJ9.eyJpZCI6IjAxOWRjMTY5LTNkZTYtNDhiZS04NzA2LTBiMWE1YzNiNmI0YSIsImRhdGEiOnt9LCJyYW5kb20iOiJjMjgzN2JkYTFjMDA0MDhmMDQyNTcwMTBhYmEzYzM0ZSJ9.enfIJaPU7GKZyCP8_Uhs1amywiqM5foSLAnZ6fooZahI8b6HFVqFovkdjdWFzcdic8wjooYvi0WP4Qw41KatcQ}{Fig. S 3} \tabularnewline 
HC$^{15}$N & $20$ & $1.3 \pm 0.3$ & $94.2 \pm 0.4$ & $4.5 \pm 0.9$ & $0.056 \pm 0.012$ & $1.0 \pm 0.5$ & \href{https://zenodo.org/records/17613203?token=eyJhbGciOiJIUzUxMiJ9.eyJpZCI6IjAxOWRjMTY5LTNkZTYtNDhiZS04NzA2LTBiMWE1YzNiNmI0YSIsImRhdGEiOnt9LCJyYW5kb20iOiJjMjgzN2JkYTFjMDA0MDhmMDQyNTcwMTBhYmEzYzM0ZSJ9.enfIJaPU7GKZyCP8_Uhs1amywiqM5foSLAnZ6fooZahI8b6HFVqFovkdjdWFzcdic8wjooYvi0WP4Qw41KatcQ}{Fig. S 3} \tabularnewline  
H$_2$CNH & $20$ & $17.8 \pm 1.5$ & $94.0 \pm 0.4$ & $9.0 \pm 0.9$ & $0.015 \pm 0.003$ & $14 \pm 7$ & \href{https://zenodo.org/records/17613203?token=eyJhbGciOiJIUzUxMiJ9.eyJpZCI6IjAxOWRjMTY5LTNkZTYtNDhiZS04NzA2LTBiMWE1YzNiNmI0YSIsImRhdGEiOnt9LCJyYW5kb20iOiJjMjgzN2JkYTFjMDA0MDhmMDQyNTcwMTBhYmEzYzM0ZSJ9.enfIJaPU7GKZyCP8_Uhs1amywiqM5foSLAnZ6fooZahI8b6HFVqFovkdjdWFzcdic8wjooYvi0WP4Qw41KatcQ}{Fig. S 7} \tabularnewline  
CH$_3$OH$^d$ & $20.5 \pm 1.8$ & $350 \pm 40$ & $94.41 \pm 0.07$ & $4.6 \pm 0.2$ & $0.32 \pm 0.04$ & $280 \pm 130$ & \href{https://zenodo.org/records/17613203?token=eyJhbGciOiJIUzUxMiJ9.eyJpZCI6IjAxOWRjMTY5LTNkZTYtNDhiZS04NzA2LTBiMWE1YzNiNmI0YSIsImRhdGEiOnt9LCJyYW5kb20iOiJjMjgzN2JkYTFjMDA0MDhmMDQyNTcwMTBhYmEzYzM0ZSJ9.enfIJaPU7GKZyCP8_Uhs1amywiqM5foSLAnZ6fooZahI8b6HFVqFovkdjdWFzcdic8wjooYvi0WP4Qw41KatcQ}{Fig. S 10} \tabularnewline 
 & $9.9 \pm 0.9$ & $380 \pm 30$ & $93.7 \pm 0.3$ & $18.9 \pm 1.1$ & $0.52 \pm 0.10$ & $310 \pm 140$ \tabularnewline
\end{longtable}

\setlength{\tabcolsep}{5pt}

\begin{longtable}[ht]{cccccccc}

\caption{\label{tab:Fits_isotopologues_molecules_on_G31_shock} Resulting physical parameters of G31.41 shock isotopologs that were not used to derive the abundances of the main isotopologs are presented. }
\\
\hline
        \hline
     Formula &  $\textit{T}_{\textrm{ex}}$ & \textit{N} & $\rm v$ & FWHM & $\tau_{max}$ & $N/N^b_{\textrm{H}_2}$ & Figure \\ 
     & (K) & ($\times$10$^{13}$ cm$^{-2}$) & (km s$^{-1})$ & (km s$^{-1})$ & & ($\times$10$^{-10}$ cm$^{-2}$)\\ \hline \endfirsthead
     \caption{Continued.}\\
\hline
     Formula &  $\textit{T}_{\textrm{ex}}$ & \textit{N} & $\rm v$ & FWHM & $\tau_{max}$ & $N/N^b_{\textrm{H}_2}$ & Figure \\ 
     & (K) & ($\times$10$^{13}$ cm$^{-2}$) & (km s$^{-1})$ & (km s$^{-1})$ & \\ \hline 
   \endhead
    \endfoot
         \hline \noalign{\smallskip}
    \multicolumn{8}{p{17cm}}{\textbf{Notes.} The values of the parameters fixed to perform some fits are shown without uncertainties. 
We also present the maximum line opacity of the transitions of each molecule ($\tau_{max}$), and the molecular abundances compared to H$_2$. The transitions used to obtain these values are in \href{https://zenodo.org/records/17613203?token=eyJhbGciOiJIUzUxMiJ9.eyJpZCI6IjAxOWRjMTY5LTNkZTYtNDhiZS04NzA2LTBiMWE1YzNiNmI0YSIsImRhdGEiOnt9LCJyYW5kb20iOiJjMjgzN2JkYTFjMDA0MDhmMDQyNTcwMTBhYmEzYzM0ZSJ9.enfIJaPU7GKZyCP8_Uhs1amywiqM5foSLAnZ6fooZahI8b6HFVqFovkdjdWFzcdic8wjooYvi0WP4Qw41KatcQ}{Table S 1}. The isotopic ratios are explained in Sect. \ref{sec:data analysis}. $^b$ $N$(H$_2$) = (1.2 $\pm$ 0.6) $\times$ 10$^{23}$ cm$^{-2}$ of G31.41 shock, more details in Sect. \ref{sec:H2_calc}.}
     \endlastfoot

H$^{13}$CN & $20$ & $8.4 \pm 0.9$ & $91.5 \pm 0.6$ & $6.9 \pm 1.7$ & $0.13 \pm 0.03$ & $7 \pm 3$
 & \href{https://zenodo.org/records/17613203?token=eyJhbGciOiJIUzUxMiJ9.eyJpZCI6IjAxOWRjMTY5LTNkZTYtNDhiZS04NzA2LTBiMWE1YzNiNmI0YSIsImRhdGEiOnt9LCJyYW5kb20iOiJjMjgzN2JkYTFjMDA0MDhmMDQyNTcwMTBhYmEzYzM0ZSJ9.enfIJaPU7GKZyCP8_Uhs1amywiqM5foSLAnZ6fooZahI8b6HFVqFovkdjdWFzcdic8wjooYvi0WP4Qw41KatcQ}{Fig. S 3} \tabularnewline \hline 
H$^{15}$NC & $20$ & $0.15 \pm 0.03$ & $94.9 \pm 0.3$ & $3.0 \pm 0.8$ & $(8.5 \pm 1.9) \times 10^{-3}$ & $0.12 \pm 0.06$ & \href{https://zenodo.org/records/17613203?token=eyJhbGciOiJIUzUxMiJ9.eyJpZCI6IjAxOWRjMTY5LTNkZTYtNDhiZS04NzA2LTBiMWE1YzNiNmI0YSIsImRhdGEiOnt9LCJyYW5kb20iOiJjMjgzN2JkYTFjMDA0MDhmMDQyNTcwMTBhYmEzYzM0ZSJ9.enfIJaPU7GKZyCP8_Uhs1amywiqM5foSLAnZ6fooZahI8b6HFVqFovkdjdWFzcdic8wjooYvi0WP4Qw41KatcQ}{Fig. S 4}\tabularnewline 
 \hline 
C$^{17}$O & $6.0 \pm 1.3$ & $(1.0 \pm 0.3) \times 10^{3}$ & $94.44 \pm 0.08$ & $2.0 \pm 0.3$ & $0.8 \pm 0.3$ & $900 \pm 500$ & \href{https://zenodo.org/records/17613203?token=eyJhbGciOiJIUzUxMiJ9.eyJpZCI6IjAxOWRjMTY5LTNkZTYtNDhiZS04NzA2LTBiMWE1YzNiNmI0YSIsImRhdGEiOnt9LCJyYW5kb20iOiJjMjgzN2JkYTFjMDA0MDhmMDQyNTcwMTBhYmEzYzM0ZSJ9.enfIJaPU7GKZyCP8_Uhs1amywiqM5foSLAnZ6fooZahI8b6HFVqFovkdjdWFzcdic8wjooYvi0WP4Qw41KatcQ}{Fig. S 5}\tabularnewline 
C$^{18}$O & $20$ & $(5.2 \pm 0.3) \times 10^{3}$ & $94.40 \pm 0.07$ & $2.35 \pm 0.14$ & $0.93 \pm 0.08$ & $4200 \pm 2000$ & \href{https://zenodo.org/records/17613203?token=eyJhbGciOiJIUzUxMiJ9.eyJpZCI6IjAxOWRjMTY5LTNkZTYtNDhiZS04NzA2LTBiMWE1YzNiNmI0YSIsImRhdGEiOnt9LCJyYW5kb20iOiJjMjgzN2JkYTFjMDA0MDhmMDQyNTcwMTBhYmEzYzM0ZSJ9.enfIJaPU7GKZyCP8_Uhs1amywiqM5foSLAnZ6fooZahI8b6HFVqFovkdjdWFzcdic8wjooYvi0WP4Qw41KatcQ}{Fig. S 5}\tabularnewline
\hline 
 
\end{longtable}

\begin{longtable}{cccccccc}
\caption{\label{tab:Poperties_molecules_on_G31_core} Results of the MADCUBA-SLIM fits of the additional molecules analyzed toward the G31.41 core,
ordered by increasing molecular mass. 
}
\\

    \hline
     Formula &  $\textit{T}_{\textrm{ex}}$ & \textit{N} & $\rm v-\rm v^c_{\rm 0}$ & FWHM & $\tau_{max}^a$ & $N/N^b_{\textrm{H}_2}$ & Figure \\ 
     & (K) & ($\times$10$^{13}$ cm$^{-2}$) & (km s$^{-1})$ & (km s$^{-1})$ & & ($\times$10$^{-9}$ cm$^{-2}$) & \\ \hline \endfirsthead

     \caption{Continued.}\\
     \hline
     Formula & $\textit{T}_{\textrm{ex}}$ & \textit{N} & $\rm v-\rm  v^c_{\rm 0}$ & FWHM & $\tau_{max}^a$ & $N/N^b_{\textrm{H}_2}$ & Figure \\ 
     & (K) & ($\times$10$^{13}$ cm$^{-2}$) & (km s$^{-1})$ & (km s$^{-1})$  & & ($\times$10$^{-9}$ cm$^{-2}$) & \\ \hline \endhead   
     
     \hline
     \endfoot

     \hline \noalign{\smallskip}
    \multicolumn{8}{p{16cm}}{\textbf{Notes.} In the case of non-detections, we indicated the upper limit in the column density with <. 
For optically thick molecules, we include the optically thin isotopolog that we used to derive the column density of the main isotopolog. The resulting physical parameters, along with their associated uncertainties, are presented. The values of the parameters that were kept fixed are shown without uncertainties.  
We also present the maximum line opacity of the transitions of each molecule ($\tau_{max}$), and the molecular abundances compared to H$_2$. The transitions of each species used to perform the fits are listed in \href{https://zenodo.org/records/17613203?token=eyJhbGciOiJIUzUxMiJ9.eyJpZCI6IjAxOWRjMTY5LTNkZTYtNDhiZS04NzA2LTBiMWE1YzNiNmI0YSIsImRhdGEiOnt9LCJyYW5kb20iOiJjMjgzN2JkYTFjMDA0MDhmMDQyNTcwMTBhYmEzYzM0ZSJ9.enfIJaPU7GKZyCP8_Uhs1amywiqM5foSLAnZ6fooZahI8b6HFVqFovkdjdWFzcdic8wjooYvi0WP4Qw41KatcQ}{Table S 2}. The isotopic ratios used and the $N$(H$_2$) used are explained in Sect. \ref{sec:data analysis}. $^a$ Maximum $\tau$ of all the transitions detected in each molecule. $^b$ $N$(H$_2$) = (1.0 $\pm$ 0.2) $\times$ 10$^{25}$ cm$^{-2}$ of the G31.41 core obtained by \cite{Mininni2020}. $^c$ v$_0$ is the systemic velocity of G31.41 which is 96.5 km s$^{-1}$ (\citealt{Beltran2018}).
    }
    
     \endlastfoot
     
\multicolumn{8}{c}{Detected molecules}  \tabularnewline \hline  
NH$_2$D & $254 \pm 22$ & $(3.8 \pm 0.3) \times 10^{3}$ & $0.99 \pm 0.18$ & $7$ & $0.128 \pm 0.018$ & $3.8 \pm 0.8$ & \href{https://zenodo.org/records/17613203?token=eyJhbGciOiJIUzUxMiJ9.eyJpZCI6IjAxOWRjMTY5LTNkZTYtNDhiZS04NzA2LTBiMWE1YzNiNmI0YSIsImRhdGEiOnt9LCJyYW5kb20iOiJjMjgzN2JkYTFjMDA0MDhmMDQyNTcwMTBhYmEzYzM0ZSJ9.enfIJaPU7GKZyCP8_Uhs1amywiqM5foSLAnZ6fooZahI8b6HFVqFovkdjdWFzcdic8wjooYvi0WP4Qw41KatcQ}{Fig. S 29} \tabularnewline 
H$_2$CNH & $254 \pm 20$ & $(1.53 \pm 0.15) \times 10^{4}$ & $0.8 \pm 0.3$ & $10.3 \pm 0.6$ & $0.072 \pm 0.012$ & $15 \pm 3$ & \href{https://zenodo.org/records/17613203?token=eyJhbGciOiJIUzUxMiJ9.eyJpZCI6IjAxOWRjMTY5LTNkZTYtNDhiZS04NzA2LTBiMWE1YzNiNmI0YSIsImRhdGEiOnt9LCJyYW5kb20iOiJjMjgzN2JkYTFjMDA0MDhmMDQyNTcwMTBhYmEzYzM0ZSJ9.enfIJaPU7GKZyCP8_Uhs1amywiqM5foSLAnZ6fooZahI8b6HFVqFovkdjdWFzcdic8wjooYvi0WP4Qw41KatcQ}{Fig. S 30} \tabularnewline 
N$_2$H$^+$ &  & $(7 \pm 3) \times 10^{3}$ &  &  &  & $7 \pm 3$ & \href{https://zenodo.org/records/17613203?token=eyJhbGciOiJIUzUxMiJ9.eyJpZCI6IjAxOWRjMTY5LTNkZTYtNDhiZS04NzA2LTBiMWE1YzNiNmI0YSIsImRhdGEiOnt9LCJyYW5kb20iOiJjMjgzN2JkYTFjMDA0MDhmMDQyNTcwMTBhYmEzYzM0ZSJ9.enfIJaPU7GKZyCP8_Uhs1amywiqM5foSLAnZ6fooZahI8b6HFVqFovkdjdWFzcdic8wjooYvi0WP4Qw41KatcQ}{Fig. S 31} \tabularnewline 
N$^{15}$NH$^+$ & $50$ & $21 \pm 4$ & $-1.6 \pm 1.1$ & $7$ & $0.08 \pm 0.04$ & $0.021 \pm 0.006$ & \href{https://zenodo.org/records/17613203?token=eyJhbGciOiJIUzUxMiJ9.eyJpZCI6IjAxOWRjMTY5LTNkZTYtNDhiZS04NzA2LTBiMWE1YzNiNmI0YSIsImRhdGEiOnt9LCJyYW5kb20iOiJjMjgzN2JkYTFjMDA0MDhmMDQyNTcwMTBhYmEzYzM0ZSJ9.enfIJaPU7GKZyCP8_Uhs1amywiqM5foSLAnZ6fooZahI8b6HFVqFovkdjdWFzcdic8wjooYvi0WP4Qw41KatcQ}{Fig. S 31} \tabularnewline
\end{longtable}

\section{Sample of molecules for the comparative study}

The complete list of molecules used in this work (detections and upper limits) is given in Table \ref{tab:Molecules_used_each_source} for the five sources used in this work (G31.41 shock, G31.41 core, L1157-B1, Sgr B2(N1) OF and G+0.693).

\begin{landscape}

\begin{longtable}{c|c|c|c|c|c|c}
\caption{\label{tab:Molecules_used_each_source} Molecules used in this work with its column densities (in cm$^{-2}$) and references ordered by increasing molecular mass.} \\
\hline
     Formula & Mol. mass & G31.41 shock & G31.41 core & L1157-B1 & Sgr B2(N1) OF & G+0.693 \\ \hline 
     \endfirsthead
    \caption{Continued.}\\
\hline
     Formula & Mol. mass & G31.41 shock & G31.41 core & L1157-B1 & Sgr B2(N1) OF & G+0.693 \\  \hline
   \endhead
    \endfoot
    \hline \noalign{\smallskip}
    \multicolumn{7}{p{21cm}}{\textbf{Notes.} Molecules with [ < ] denotes non detections (upper limits) and [ - ] denotes that the molecule has not been identified. \textbf{References.} a) This work; b) \cite{Fontani2024}; c) \cite{Mininni2020}; d) \cite{Colzi2021}; e) \cite{Mininni2023}; f) \cite{Lopez-Gallifa2024}; g) \cite{Garcia-de-la-concepcion-2022}; h) \cite{Lefloch2021}; i) \cite{Sugimura2011}; j) \cite{Lefloch2017}; k) \cite{Lefloch2016}; l) \cite{Holdship2019}, m) \cite{Gomez-Ruiz2015}; n) Recalculated from ASAI data assuming 18$''$of source size which is concordant with \cite{Lefloch2017}; o) \cite{Podio2014}; p) \cite{Busch2024}; r) \cite{Mendoza2014}; s)\cite{Mendoza2018}; t) Isotopologs from \cite{Colzi2022} and then isotopic ratio used of $^{14}$N/$^{15}$N; u) \cite{Zeng2018}; v) \cite{Rodriguez-almeida2021a}; w) \cite{Requena-Torres2008}; x) \cite{Sanz-novo2024b}; y) \cite{Sanz-novo2022a}; z) Private communication with Dr. Rivilla; aa) \cite{Rivilla2018}; ab) \cite{Sanz-novo2024a}; ac) \cite{Jimenez-Serra2022}; ad) Sanz-novo in prep 2025; ae) \cite{Rivilla2020b}; af) \cite{Rivilla2022b}; ag) \cite{Bizzocchi2018}; ah) \cite{Rivilla2023}; ai) \cite{Zeng2023}; aj) \cite{Rivilla2022a}; ak)\cite{Sanz-novo2023}; al) \cite{SanAndres2024}; am) \cite{Rivilla2021a}.} 
     \endlastfoot
\hline
HCN & 27 & $(4.4 \pm 1.4) \cdot 10^{15}$ $^a$ & $(1.5 \pm 0.4) \times 10^{18}$ $^f$ & $6.6_{-0.6}^{+1.1} \times 10^{14}$ $^h$ & - & $(4.6 \pm 1.2) \times 10^{16}$ $^t$\tabularnewline 
H$_2$CNH & 29 & $(1.80 \pm 0.15) \cdot 10^{14}$ $^a$ & $(1.53 \pm 0.15) \cdot 10^{17}$ $^a$ & - & - & $(5.4 \pm 0.3) \times 10^{14}$ $^u$ \tabularnewline  
CH$_3$NH$_2$ & 31 & $< 1.4 \cdot 10^{14}$ $^a$ & $(1.6 \pm 0.3) \cdot 10^{17}$ $^a$ & - & - & $(3.0 \pm 0.9) \times 10^{13}$ $^u$\tabularnewline 
CH$_3$OH & 32 & $(3.5 \pm 0.4) \times 10^{15}$ $^a$ & $(7 \pm 3) \times 10^{19}$ $^e$ & $(1.00 \pm 0.10) \times 10^{15}$ $^j$ & $(3.5 \pm 0.8) \times 10^{17}$ $^p$ & $(1.5 \pm 0.3) \times 10^{16}$ $^v$\tabularnewline 
CH$_3$CN & 41 & $(2.42 \pm 0.13) \cdot 10^{14}$ $^a$ & $(2.7 \pm 0.4) \times 10^{17}$ $^e$ & $(7.7 \pm 0.3) \times 10^{12}$ $^n$ & $(2.8 \pm 0.5) \times 10^{17}$ $^p$ & $(1.15 \pm 0.03) \times 10^{14}$ $^u$ \tabularnewline 
CH$_3$NC & 41 & $< 8 \cdot 10^{11}$ $^a$ & $(1.1 \pm 0.3) \times 10^{15}$ $^f$ & - & - & $<3 \times 10^{12}$ $^u$ \tabularnewline 
NH$_2$CN & 42 & $(5.0 \pm 0.5) \cdot 10^{12}$ $^a$ & $(1.11 \pm 0.08) \times 10^{15}$ $^f$ & - & $< 4 \times 10^{14}$ $^p$ & $(3.1 \pm 0.2) \times 10^{14}$ $^u$ \tabularnewline 
H$_2$CCO & 42 & $(7.1 \pm 0.8) \cdot 10^{13}$ $^a$ & $(6.0 \pm 0.5) \cdot 10^{16}$ $^a$ & $(1.7 \pm 0.7) \times 10^{13}$ $^j$ & - & $3 \times 10^{14}$ $^w$ \tabularnewline
\end{longtable}

\end{landscape}

\section{Comparative plots for O-, N-, and S-/P-bearing species}

\begin{figure}[ht]
\includegraphics[width=0.9\textwidth]{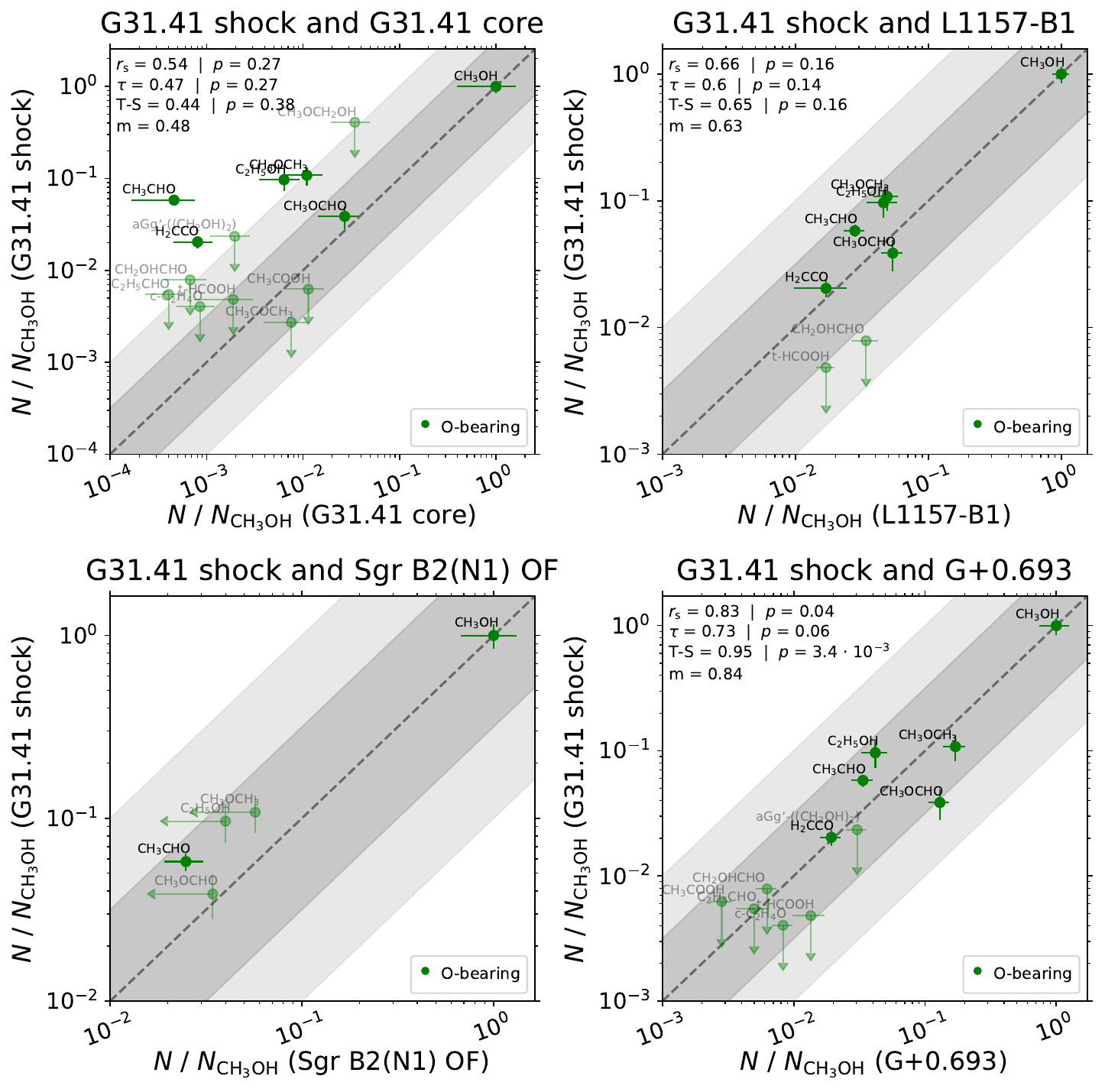}
\caption{Comparison of the molecular ratios of the O-bearing molecules with respect to \ch{CH$_3$OH} in the G31.41 shock with those in G31.41 core (\textit{upper left}), L1157-B1 shock (\textit{upper right}), Sgr B2(N1) OF shock (\textit{lower left}), and G+0.693 molecular cloud (\textit{lower right}). The upper limits for undetected species are indicated with arrows. The light and dark shaded gray area corresponds to half and one order of magnitude scatter from the y=x relation denoted by the dashed line. The results of the correlation tests are indicated in the upper left corner of each panel: $r_{\mathrm{s}}$ (Spearman), $\tau$ (Kendall), and T-S (Theil-Sen), with their associated $p$-values, and the average $m$ (for details see Sect. \ref{sec:Comparison of molecular abundances}).}
\label{fig:Correlaciones_fuentes_O}
\end{figure}

\begin{figure*}
\includegraphics[width=0.9\textwidth]{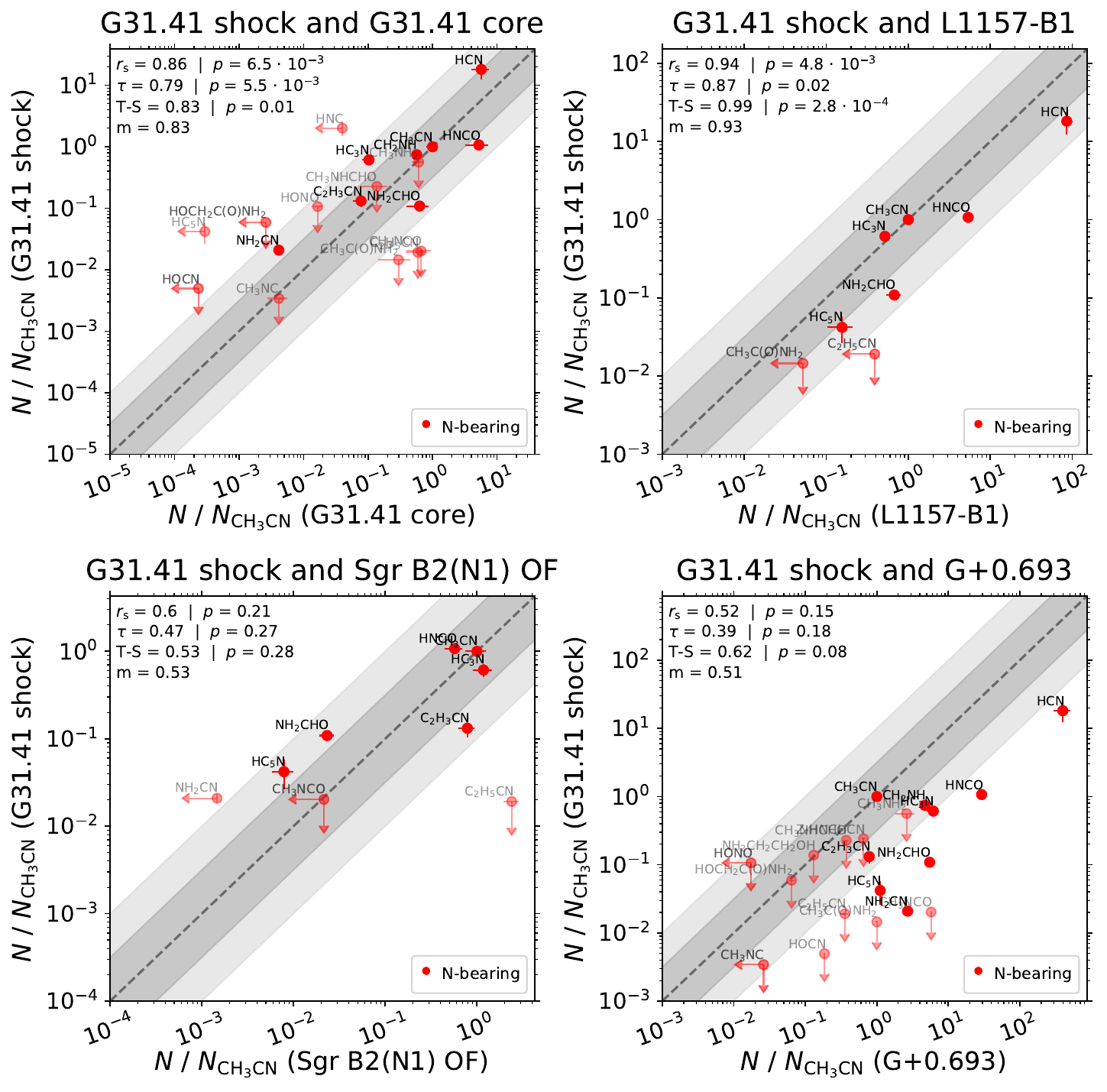}
\caption{Comparison of the molecular ratios of the N-bearing molecules with respect to \ch{CH$_3$CN} in the G31.41 shock with those in G31.41 core (\textit{upper left}), L1157-B1 shock (\textit{upper right}), Sgr B2(N1) OF shock (\textit{lower left}), and G+0.693 molecular cloud (\textit{lower right}). The upper limits for undetected species are indicated with arrows. The light and dark shaded gray area corresponds to half and one order of magnitude scatter from the y=x relation denoted by the dashed line. The results of the correlation tests are indicated in the upper left corner of each panel: $r_{\mathrm{s}}$ (Spearman), $\tau$ (Kendall), and T-S (Theil-Sen), with their associated $p$-values, and the average $m$ (for details see Sect. \ref{sec:Comparison of molecular abundances}).}
\label{fig:Correlaciones_fuentes_N}
\end{figure*}

\begin{figure*}
\includegraphics[width=0.9\textwidth]{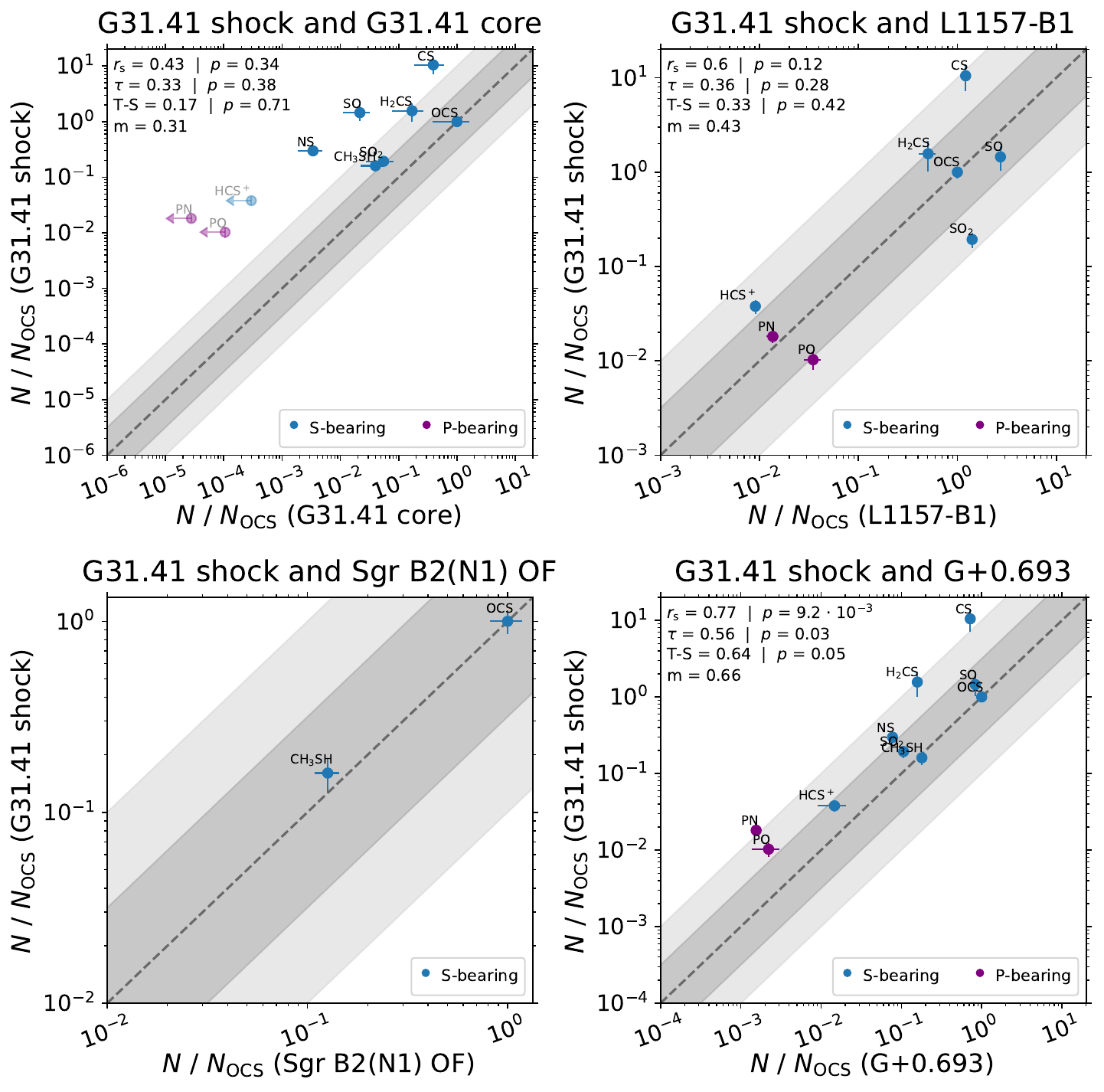}
\caption{Comparison of the molecular ratios of the S-/P-bearing molecules with respect to \ch{OCS} in the G31.41 shock with those in G31.41 core (\textit{upper left}), L1157-B1 shock (\textit{upper right}), Sgr B2(N1) OF shock (\textit{lower left}), and G+0.693 molecular cloud (\textit{lower right}). The upper limits for undetected species are indicated with arrows. The light and dark shaded gray area corresponds to half and one order of magnitude scatter from the y=x relation denoted by the dashed line. The results of the correlation tests are indicated in the upper left corner of each panel: $r_{\mathrm{s}}$ (Spearman), $\tau$ (Kendall), and T-S (Theil-Sen), with their associated $p$-values, and the average $m$ (for details see Sect. \ref{sec:Comparison of molecular abundances}).}
\label{fig:Correlaciones_fuentes_S}
\end{figure*}

Figures \ref{fig:Correlaciones_fuentes_O}, \ref{fig:Correlaciones_fuentes_N}, and  \ref{fig:Correlaciones_fuentes_S} are the same plot as Fig. \ref{fig:Correlaciones_fuentes_colores}, but each plots corresponds to O-, N- and S-/P- bearing family respectively.

\end{document}